\documentclass[12pt,preprint]{aastex}

\shorttitle{Observational Comparison between ULXs and XRBs}
\shortauthors{Berghea et al.}

\begin{document}

\title{Testing the Paradigm that Ultraluminous X-ray Sources as a Class 
Represent Accreting Intermediate-Mass Black Holes}

\author{C. T. Berghea}
\affil{Department of Physics, The Catholic University of America, Washington, DC 20064}
\email{79berghea@cua.edu}

\author{K. A. Weaver}
\affil{Laboratory for High Energy Astrophysics, NASA's Goddard Space Flight Center, Greenbelt, MD 20771}
\email{kweaver@milkyway.gsfc.nasa.gov}

\author{E. J. M. Colbert}
\affil{Department of Physics and Astronomy, Johns Hopkins University, Baltimore, MD 21218}
\email{colbert@jhu.edu}
\and

\author{T. P. Roberts}
\affil{Department of Physics, Durham University, South Road, Durham DH1 3LE, UK}
\email{t.p.roberts@durham.ac.uk}

\begin{abstract}

To test the idea that ultraluminous X-ray sources (ULXs) in external 
galaxies represent a class of accreting intermediate-mass black holes 
(IMBHs), we have undertaken a program to identify ULXs and a lower 
luminosity X-ray comparison sample with the highest quality data in the 
{\it Chandra} archive. We establish as a general property of ULXs that the 
most X-ray-luminous objects possess the flattest X-ray spectra (in the {\it Chandra} bandpass). 
No prior sample studies have established the general hardening of ULX spectra 
with luminosity. This hardening occurs at the highest luminosities 
(absorbed luminosity $\geq$5$\times$10$^{39}$~erg~s$^{-1}$)
and is in line with recent models arguing that ULXs are 
actually stellar-mass black holes. From spectral modeling, we show that 
the evidence originally taken to mean that ULXs are IMBHs - i.e., the 
``simple IMBH model'' - is nowhere near as compelling when a large sample 
of ULXs is looked at properly. 
During the last couple of years, {\it XMM-Newton}
spectroscopy of ULXs has to a large extent begun to negate the simple 
IMBH model based on fewer objects. We confirm and expand these results, 
which validates the {\it XMM-Newton} work in a broader sense with independent X-ray 
data. We find that (1) cool-disk components are present with roughly 
equal probability and total flux fraction for any given ULX, regardless 
of luminosity, and (2) cool-disk components extend below the 
standard ULX luminosity cutoff of 10$^{39}$~erg~s$^{-1}$, 
down to our sample limit of 10$^{38.3}$~erg~s$^{-1}$.
The fact that cool disk components are not correlated with luminosity 
damages the argument that cool disks indicate IMBHs in ULXs, for which 
strong statistical support was never found. 

\end{abstract}


\section{INTRODUCTION}

Ultraluminous X-ray sources (ULXs) have long been hailed as direct
observational evidence for the existence of accreting
intermediate-mass black holes \citep[IMBHs;][]{col99}. The X-ray
spectral model that has emerged as a central pillar for this argument
(the ``simple IMBH model'') is that which is commonly applied
as the canonical X-ray spectral fit to Galactic black-hole binaries
with stellar mass black holes \citep{mcc06}.  This
model consists of a thermal accretion disk component plus a power-law (PL)
continuum component. When applied to ULX spectra, the derived disk
temperatures are 0.1$-$0.3~keV \citep[e.g.][]{mill04}, much lower
than Galactic black holes (at 0.6$-$1~keV). A cooler disk implies a
bigger disk; so assuming that the disk approximately extends inward to the
last stable orbit around the black hole, this would imply bigger, and
more massive black holes. Such cool disks were indeed found in a few ULXs \citep[e.g.][]{mill04}.

To counter this argument, many recent papers have pointed out both
theoretical and observational problems with the simple IMBH model as a
global explanation for all ULXs \citep[e.g.][]{gon06, wil06, rob07}.
The observed accretion disk components can be fairly
weak; thus they do not provide a reliable measure of black hole
mass. Also, the simple IMBH model does not necessarily approximate
well the X-ray spectra of many ULXs. Attention has switched to perhaps
less exotic models to explain some of the ULXs, such as beaming
\citep{king01} or super-Eddington accretion \citep{beg02}, both of
which explain ULX X-ray properties without the need for an
IMBH. Galactic super-Eddington sources are known, such as stellar-mass
black hole binaries like GRS~1915+105 \citep{fen04}, V4641~Sgr
\citep{rev02} and possibly SS~433. The latter could be an example of
both beaming {\it and} super-Eddington emission, the combination of
which could easily explain even the most luminous ULXs \citep{beg06, pou07}.
Cool accretion disks can also be physically explained by ``coupled disk-corona'' models
\citet{don06}, blurred emission and absorption lines from surrounding
(outflowing) gas \citep{gon06}, or a microblazar with magnetized jets \citep{fre06},
that can transfer disk energy into the jet (thus making the disk
fainter and cooling it the same time).

Recent detailed X-ray spectral modeling has revealed properties that
further complicate any simple global interpretation, suggesting that
multiple classes of ULXs exist.  Some very bright ULXs have been found
by several authors \citep{zez02, sor07, sorw06, soc06} to have relatively
flat spectra, not usually expected in high states for accretion states
of black holes \citep{mcc06}. A flat spectrum suggests an inverse
correlation between the slope of the spectrum and source luminosity
\citep[see also NGC~5204 X-1;][]{rob06}. Such an inverse correlation
is hard to explain with current IMBH models, because in the typical
high state the spectrum is soft, dominated by the disk component and
with a steep PL \citep{mcc06}. Specifically, {\it XMM-Newton} spectroscopy
of ULXs has to a large extent already begun to directly negate the
simple IMBH model \citep[see, e.g.][]{wil06, gon06}.

Two {\it Chandra} surveys suggest that ULXs may in fact be an extension 
of normal lower-luminosity galaxy populations to higher luminosities. 
Using simple PL models applied to spectra with typically ~50 
counts each, \citet{swa04} compared ULXs to a lower luminosity sample of 
X-ray sources with L$_{X}=$~10$^{38}-$10$^{39}$~erg~s$^{-1}$ and found 
both distributions of photon indices to be well fitted by Gaussians 
centered at about 1.9. The samples also have similar X-ray colors, time 
series and positions within their host galaxies. In another {\it Chandra} study, 
\citet{col04} found no discernible difference between the 
X-ray colors of ULXs and lower luminosity sources in spiral galaxies. 
Both analyses were done with data of fairly poor spectral quality in 
terms of fitting detailed models and the latter relied on a color-color 
analysis rather than spectral fitting. These works also did not 
include two-component model fits that would identify spectral states 
and directly test the simple IMBH model

To search carefully for spectral properties that can differentiate
ULXs, we need the best data available that will allow us to
distinguish between simple spectral models. In this paper we use the
highest quality X-ray spectra for a large, complete sample of ULXs
from the {\it Chandra} archives to test various ULX models.  We are able to
provide a statistically strong comparison of the results with
lower luminosity X-ray sources of equal data quality.  We pay special
attention to properties usually associated with ULXs, such as the
signature of a cool disk, which has never been tested for a uniform
and large sample of good quality {\it Chandra} spectra. Is the cool
disk preferentially found in ULXs? If yes, does the disk dominate the
total emission? We also search for other spectral behaviors found more
recently in individual ULXs, such as a correlation between hardness
and luminosity, and what this might mean.  For the first time, strong
statistical tests of various ideas of ULX models can be provided to
the ULX community.

With its unmatched spatial resolution, {\it Chandra} is better suited
than {\it XMM-Newton} for studying point sources in crowded regions
or resolving point sources in distant galaxies.  This is particularly
true for the starburst galaxies that host populations of ULXs (e.g NGC
3256, Lira et al. 2002; Cartwheel galaxy, Gao et al. 2003), where only
{\it Chandra}'s unparalleled X-ray optics can spatially and spectrally
resolve the emission of ULXs from that of the underlying galaxy. We
have searched all public data available in the {\it Chandra} archive
for ULXs and lower-luminosity comparison objects with at least 1000
counts. In section 2 we present our source selection process, methods
for identifying rejected objects, and an estimate of contamination
from background objects. In section 3 we discuss the spectral fitting
procedures and compare the spectral properties of the two samples. Our
goal is to determine whether ULXs as a class have different spectral
properties than the less luminous, ``normal'' X-ray sources and to
offer an improved diagnosis by using the high-quality spectral data
available in the {\it Chandra} archive. In section 4 we present
results from the variability analysis. Finally, in Section 5
we interpret our results and discuss the insight provided into the
nature of ULXs.

\section{SAMPLE SELECTION AND OBSERVATIONS}

\subsection{Sample Criteria}

There are many published papers that address the nature of ULXs. These 
analyses are typically drawn from heterogeneously selected samples, small
numbers of objects, or large samples with limited data quality 
\citep{swa04, col04}. Comparisons of ULXs with 
other types of X-ray point sources in nearby galaxies often use 
selection criteria that do not provide the spectral data quality  
that allows a robust set of statistical conclusions 
to be drawn from data modeling. In this work, we use
criteria that create the best possible available sample to address 
the nature of ULXs by defining a large and statistically robust sample of
ULXs and other pointlike X-ray sources in nearby galaxies with uniform data quality. 
Uniformity of data quality is our prime objective, and the completeness of
our sample is limited by the observations that are available in the 
{\it Chandra} archive, most of which have been obtained by other researchers for a variety of purposes.

Our ULX and comparison samples are statistically robust in the sense that we include
all point sources in the {\it Chandra} archive with at least 1000 counts and a
luminosity above 10$^{38.3}$~erg~s$^{-1}$. We carefully reject sources 
associated with active galactic nuclei, supernovae, and foreground stars. We 
also reject piled-up observations to simplify our spectral analysis. 
{\it Chandra} provides the most accurate X-ray positions to date; thus
we can be sure to identify well-isolated objects for
our study. Several {\it XMM-Newton} studies of ULXs are published, but while 
these individual spectra are of higher quality, there are fewer individual point sources 
available due to {\it XMM-Newton's} poorer imaging resolution and source confusion 
for faint targets located in crowded regions in galaxies. We have not 
made our sample fully representative in the sense of picking the same
number of ULXs and comparison objects from similar galaxy types. Conclusions 
about the distribution of objects according to galaxy type can, however, be 
inferred from our statistical comparisons. Uniformly good-quality 
X-ray spectra allow us to apply exactly the same physical models to the 
ULX and comparison samples and directly compare results
within the sensitivity limits.

There are selection biases inherent in our analysis. One is distance. 
For sources that are intrinsically less luminous, a larger fraction of objects 
will be located in the nearest host galaxies,
while more luminous objects can be utilized from galaxies at greater
distances. We also do not select objects according to any specific 
requirement of their local environments (e.g., their locations in their host galaxies).

\subsection{Initial Sample}

Our sample is derived from the list of X-ray point sources generated by
the XASSIST\footnote{XASSIST (Ptak \& Griffiths 2003)
is a semiautomatic X-ray analysis program written and maintained by
A.~Ptak. Analysis of archival data processed by XASSIST can
be found at http://www.xassist.org} {\it Chandra} pipeline. 
For manageability, we have chosen all {\it Chandra} ACIS sources in the public 
archives as of a cutoff date of October 18, 2004. We determine 
which XASSIST sources are associated with host galaxies following the 
procedure used by \citet{col02}. X-ray sources are further considered 
if they are located inside the D$_{25}$ ellipse of their
host galaxy. Parameters for the D$_{25}$ ellipse 
are obtained from v3.9b of the Third Reference Catalog of Bright Galaxies 
\citep[RC3;][]{dev91}. We further only consider RC3 galaxies with
recessional velocities $cz \le$ 5000 km~s$^{-1}$. 

To estimate observed X-ray luminosities L$_{XA}$,
we calculated the 0.3$-$8.0 keV fluxes with XASSIST assuming a PL 
model with $\Gamma=$~1.8 and Galactic absorption and used distances 
for the associated RC3 galaxies. For galaxies with $cz <$ 1000 km~s$^{-1}$, 
distances were taken from \citet{tul88}, otherwise
distances were computed using H$_0$~$=$~75~km~s$^{-1}$~Mpc$^{-1}$.
We retained all sources with L$_{XA}~>$~10$^{38.3}$~erg~s$^{-1}$
and then manually inspected the X-ray images 
to eliminate false X-ray sources chosen by the automatic data 
processing. This initial selection yielded 126 unique X-ray point 
sources in 188 {\it Chandra} observations. A significant fraction 
of the point sources were observed multiple times, which provides
some useful variability information.

\subsection{Obvious Rejected objects (AGNs, QSOs, SNe, Stars \& Jets)}

To reject X-ray point sources unrelated to our science, 
we used the NASA/IPAC Extragalactic Database (NED)\footnote{Available at http://nedwww.ipac.caltech.edu}. 
The absolute positional uncertainty for {\it Chandra} ACIS images is better than 
1$\arcsec$ \citep[e.g.][]{wei03}, which provides the accuracy required 
to identify an optical, infrared, or radio counterpart. 
Optical positions provided by NED are typically accurate to 
within a few arc-seconds, and positions may be slightly less accurate 
for infrared and radio sources, so we first searched NED using 
a radius of 5$\arcsec$ 
surrounding the XASSIST position. We next visually inspected the 
X-ray sources and their possible NED counterparts by overlaying the 
XASSIST position, the NED position, and the D$_{25}$ galaxy ellipse 
onto the raw X-ray images and DSS2\footnote{The Second Digitized 
Sky Survey consists of high-resolution scans of several plate collections 
in the red, blue, visible, and near-infrared. 
The images were downloaded from the server installed at ESO, using a remote 
client, the ESO/ST-ECF Digitized Sky Survey application.} 
red images. 

Optical images were used to check for bright, foreground 
star counterparts. We then refined our identification search by 
examining the literature for more accurate positions 
for identified NED counterparts. In some cases 
VLBI measurements are available with sub-milliarcseconds positional 
accuracy, such as those used by \citet{ma98} for the International 
Celestial Reference Frame (ICRF). Some published 2MASS positions also 
use the ICRF reference system and have accuracies 
better than 0.1$\arcsec$, varying slightly with the source brightness 
\citep[see UCAC2;][]{zac04}.
Optical positions that can be correlated with 
radio measurements show systematic differences of only 0.1$\arcsec$ 
\citep[e.g.][]{arg90}. Overall, we determined that the 
positional uncertainties of identified counterparts are generally much 
smaller than our X-ray positional uncertainties, the largest 
uncertainty being 1$\arcsec$. We estimated a conservative upper 
limit of 1.5$\arcsec$ for the net uncertainty in separation 
between the {\it Chandra} X-ray source and an identified optical, IR, or radio 
counterpart for any object in our sample. Therefore, we feel 
confident that we have identified correct 
counterparts to within the errors provided by the X-ray data.

From this search we reject 32 X-ray sources out of 69 {\it Chandra} observations.
Most are associated with Seyfert and LINER galaxies \citep{ho97, ver03, bry99}.
Others include background quasars and pointlike X-ray 
knots associated with jets within the host galaxy. As an example, 
source 37 in \citet{zez02}, in the Antennae galaxy pair (NGC~4038/4039), 
is a background quasar with redshift 0.26 \citep{cla05}.
We identified a supernova in NGC~891 \citep[SN~1986J;][]{bie02}. 
One ULX in M101 \citep[NGC 5457 X-6;][]{rob2000} is actually 
a foreground star, GSC 2.2~3842. After rejecting sources based
on optical and NED counterparts we are left with 
94 X-ray sources in 119 {\it Chandra} observations.

\subsection{Reprocessing of Archival Data and Final Rejection Criteria}

Having narrowed our sample according to the above criteria, the 
ACIS imaging data were retrieved from the {\it Chandra} archives.  
The level-1 event files were reprocessed with CIAO
v3.0.1 and CALDB v1.4, using the {\sc acis\_process\_events} tool.
No adjustment was made for charge transfer inefficiency (CTI) effects
between pixels during the data readout. This allows the analysis of 
data to be uniform for different CCD detectors and degrees of CCD pile-up.
To minimize pileup effects, we restricted the count rates for on-axis 
full-frame (frame time 3.24~s) CCD observations to be   
$<$0.08~s$^{-1}$. According to the {\it Chandra} 
Proposer's Observatory Guide\footnote{Available at http://cxc.harvard.edu/proposer/POG/}, 
this corresponds to 10\% pileup. Count rates in excess of this 
value for point sources are likely to impact the extracted spectra. 
PHA randomization was applied, but pixel randomization was not.

In cases where the X-ray sources were observed off-axis or in a subarray CCD mode,
the pileup effect is reduced, and we can accept a higher net count rate. 
The actual pile-up fraction is estimated for the ``reduced'' count rates in 
Table~\ref{table1}. These count rates were calculated by taking into account the larger 
point-spread function for sources observed off-axis and the CCD observation mode.
Column (8) of the table lists the ACIS CCD in which the source is imaged and the subarray value,
i.e., the fraction of the CCD used in the observation. Exposing a smaller chip 
area results in shorter frame times and reduces the pileup.

Source spectra were typically extracted from regions of radius 2$\arcsec$, 
and local background spectra were extracted using annuli 
with inner and  outer radii of 6$\arcsec$ and 10$\arcsec$. 
For off-axis sources we used elliptical regions, 
and for crowded regions, slightly more complicated background 
regions, as needed. Visual inspection ensured 
that there was no confusion with any nearby X-ray sources.
Sources were retained that had $>$1000 counts in the reprocessed data.
Spectral fitting was performed using XSPEC v11.2.0bd.

A total of 21 sources had less than 1000 counts after the archival data 
were fully reprocessed, so these are rejected. In addition, 
9 observations of 7 sources have $>$10\% pile-up and are rejected 
(Table~\ref{table1}). Most of the sources with significant pile-up have
other {\it Chandra} observations, so only 3 unique objects are 
fully rejected from our sample because of pile-up, 
2 ULXs and one lower luminosity source.

\subsection{Final Sample}

A total of 69 unique objects in 89 data sets comprise our final sample.
The properties of these objects are listed in Table~\ref{table2}, 
together with some properties of their host galaxies.
Using count rates derived from our reprocessed data, we re-computed
the 0.3$-$8.0~keV observed luminosities (L$_X$), using a
PL model with $\Gamma=$~1.8 and Galactic absorption. A final 
division into two groups is made according to the 
maximum observed luminosity, L$_X^{max}$. There are 47 ULXs 
(L$_X^{max}$~$\ge$~10$^{39.0}$ erg~s$^{-1}$)
and 22 comparison objects of lower-luminosity
(L$_X^{max}$~$<$~10$^{39.0}$ erg~s$^{-1}$).

Some sources show luminosity variability. 
For two ULXs, U2 (M33~X-8) and U41 (IXO~83), their luminosity can fall below 
our threshold value of 10$^{39.0}$ erg~s$^{-1}$ in some cases, but we still
retain the classification of ULX.
Our method identifies a ULX as such if it is observed with 
L$_X$~$\ge$~10$^{39.0}$ erg~s$^{-1}$ at least once.
On the other hand, a well-known and previously studied ULX, IXO~85 
(C22) is excluded from our ULX sample because the {\it Chandra} 
luminosity falls just below our ULX limit.

Examining the galaxy properties in Table~\ref{table2}, we find that most of 
our sample objects reside in 
spiral or irregular (merger) galaxies and are preferentially located
in spiral arms and star-forming regions. Our galaxy sample includes two 
mergers (NGC 520 and the Antennae) and four early type galaxies 
(NGC 2681, NGC 4125, M87, and Cen A).
We see little difference between the ULX locations in their host galaxies
in general and the locations of the comparison sources. The two groups
also tend to have similar deprojected offsets from the centers of their galaxies.
Two ULXs (U2 and U14) are associated with 
the nucleus of their host galaxies (M33 and NGC 3310, respectively), 
but with no evidence of AGN activity. We do not have enough detailed 
information on these sources to know what fraction are known low-mass X-ray 
binaries (LMXBs) and high-mass X-ray binaries (HMXBs). 
The identification of the optical counterparts would require sensitive 
optical imaging. Based on their location in the host galaxy, we can only say that 
most of our sources in both samples are consistent with being HMXBs.

\subsection{Background Contamination}

Here we estimate potential sample contamination from additional 
background objects that have not already been clearly identified.  
We use the log(N)$-$log(S) function from X-ray deep field surveys to 
estimate the fraction of additional background objects based on our 
source fluxes and galaxy distances. For our 
sample criteria, we construct two flux limits, FL and FC. FL is the flux of 
a source with a specific luminosity: 10$^{39.0}$~erg~s$^{-1}$ for ULXs 
and 10$^{38.3}$~erg~s$^{-1}$ for our comparison sample. FC is the flux 
of a source that provides 1000 counts in its spectrum for the longest 
exposure time obtained for each galaxy. Assuming a PL model with 
$\Gamma=$~1.8 and using the Galactic value of absorption corresponding
to the location of the center of each galaxy on the sky we calculate 
FL and FC for all of the 286 galaxies in our original list 
(see section 2.1). The final flux limit for each galaxy to compare 
with log(N)$-$log(S) is the largest of the two fluxes, FC or FL. 
All of our measured fluxes are above 
10$^{-14}$~erg~cm$^2$~s$^{-1}$, which corresponds to an ACIS count rate 
of $\sim$10$^{-3}$~s$^{-1}$. 

To make a background estimate we also need to account for the size
of the detectors on the sky compared to the projected sizes of the galaxies.
The area of each galaxy in deg$^2$ is first calculated within the D$_{25}$ ellipse. 
Most observations are done in ACIS imaging mode, 
with detector areas of $\sim$0.117~deg$^2$ for both ACIS-I and ACIS-S.
Data can be extracted from specific CCD chips, and some observations are only 
in subarray mode with a significantly smaller exposed area. The disparity 
between the sizes of the galaxies and the detector coverage can 
affect our background estimates. For the nearest galaxies, 
their size on the sky is larger than or comparable to the size of the ACIS detectors.
Naturally, if the projected area of the galaxy is larger than or 
comparable to the size of the detector, these galaxies will
provide the largest estimated contributions to the background counts.
We therefore account for the fractional coverage of the 13 largest galaxies by 
over plotting the CCDs and estimating the coverage fraction. These 13 galaxies 
(out of the original 286) contribute 65\% to the total estimate 
of the contamination. For the remaining galaxies we use the D$_{25}$ ellipse area. 

We used the log(N)$-$log(S) function from two separate surveys to obtain 
flux estimates. The popular {\it ROSAT} deep survey 
in the Lockman Hole \citep{has98} gives log(N)$-$log(S) 
for the flux interval 10$^{-15}-$10$^{-13}$~erg~cm$^2$~s$^{-1}$, 
in the range 0.5$-$2~keV. We apply a scale factor of 0.38 for our 
0.3$-$8~keV band, obtained using the absorbed PL model with $\Gamma=$~1.8.
The {\it Chandra} Multiwavelength Project (ChaMP) serendipitous survey 
\citep{kim04} contains a larger sample, and covers a wide area 
($\sim$14 deg$^2$). It uses the same soft X-ray band as the {\it ROSAT} deep survey,
but the slope of the log(N)$-$log(S) function is shallower at the 
high end.

For ULXs, the {\it ROSAT} and ChaMP surveys predict no more than 3 or 
5 spurious sources, respectively.
For our lower luminosity objects, the prediction is 1 or 2 spurious sources.
The survey estimates are compatible given large errors due to 
poor sampling at the high flux end. Thus, no more than 
approximately one in ten sources in our sample is likely a background object. 
In a practical sense this is an upper limit, as our estimate does not 
take into account the variable absorption column through each galaxy, which will 
attenuate the signal of any background sources shining through the galaxy 
(i.e. reduce their observed flux). This is especially important, 
as we have used surveys in the 0.5$-$2 keV band where 
absorption is strong. We also remind the reader that we have already 
identified and rejected two background quasars (Section~2.3).

\section{SPECTRAL ANALYSIS}

We grouped the spectra to have a minimum of 15 counts per energy bin
for the energy range of 0.3$-$8.0~keV.
All fits were performed using the Galactic absorbing column
(as listed in Table~\ref{table2}), plus an intrinsic absorbing column 
for each galaxy.
Galactic values were obtained with the COLDEN routine in CIAO,
which provides a foreground N$_H$ value at a given celestial position.
We chose to define acceptable (or ``good'') fits as
those for which $\chi_{\nu}^{2}\leq~$1.2. Unless specified, all errors quoted
are 90\% confidence for one interesting parameter ($\Delta\chi^{2} =$~2.7).
For sources with multiple observations, the individual observations were
first fitted separately and then all observations were fitted together
in XSPEC for the various purposes of our work. Simultaneous fits
are used in the histograms and listed in the tables (e.g. Table~\ref{table3})
and individual fits are shown in some of the plots to demonstrate
any variability in luminosity and spectral shape. For the simultaneous
fits, the model parameters were constrained to the same value in XSPEC,
and only the normalizations of model components were allowed to vary freely.

In a statistical sense, spectral fitting results can be strongly biased 
by the number of counts in each spectrum. To test for such biases between
the ULX and the comparison samples, we constructed histograms of
net counts in the spectra (see Figure~1). For sources with multiple observations, 
we chose the observation that contained the largest number of
counts in its spectrum (see Table~\ref{table2}) to represent in the
histogram. The distributions of the number of counts for objects in the
samples are similar. A Kolmogorov-Smirnov (K-S) test provides a
probability of 0.86, indicating that we have no reason to reject the
hypothesis that the distributions are identical in count space. Thus,
the two samples have equal sensitivity to spectral features for our model 
fitting.

\subsection{Single-Component Spectral Fits}

Recent spectral analysis of {\it Chandra} ULX spectra shows that many 
are well fitted by simple models \citep[e.g.][]{hum03,swa04}.
We therefore fit all the spectra with either an absorbed
PL model or a multi-color disk blackbody (MCD) model,
with absorption fixed at the Galactic value in XSPEC.
To keep our results within physical bounds, we impose upper limits of
$\Gamma\leq$~10 and kT$_{in}\leq$~4~keV, respectively.
The results are listed in Table~\ref{table3}.
Both the ULX and the comparison samples are generally well
fitted by the absorbed PL model (66\% and 50\%, respectively, are good fits, 
as indicated in col. [5] of the table). 
For the absorbed MCD model, good fits comprise 45\% and 50\% of the samples, respectively.

The histograms in Figure~2 show the distributions of the photon index
and inner disk temperature, normalized to allow for easy comparison. 
For the full sample, we find no significant difference between
ULXs and lower luminosity objects. Luminosity dependences are
presented in Figure~3. For objects that have multiple observations,
all fit results are shown.

We have applied the K-S test and
the T-test to the samples in different ways.
The first row of Table~\ref{table4} shows the results of the test applied to the total set of fits,
while the second row is restricted to the ``good'' fits as defined
in the first paragraph of Section~3. All calculated probabilities
are higher than a 5\% significance level, confirming that there
are no significant differences when comparing the distributions
or their means. We note that the derived probabilities differ in
some cases significantly between the K-S test and the T-test,
which is an indication that the distributions plotted in Figure~2
are possibly derived from intrinsic samples that do not have
normal distributions and/or that our sample sizes are small (such
tests are usually more reliable when applied to large samples).

Even with these caveats, we find an interesting trend if we
limit our sample further. When only considering the good fits,
the disk temperatures are marginally higher for ULXs 
(at 1.8~keV, with a significance level of 7\%-8\%). 
If we further use Figure~3a to split the ULXs
themselves into two groups, with a luminosity break at
5$\times$10$^{39}$~erg~s$^{-1}$, then we find that the
highest luminosity ULXs have significantly harder X-ray spectra
than both the lower luminosity ULXs and the comparison sample
(rows 3-6 of Table~\ref{table4}).

Our primary result from applying single-component models is that
all of the highest luminosity ULXs that are well fitted by the PL
model possess hard X-ray spectra ($\Gamma\leq$~2 and
kT$_{in}\geq$~1.3~keV). The most luminous ULXs have harder spectra,
and those that are less luminous have spectral shapes similar to the
comparison sample. Not all of the high-luminosity ULXs have
hard spectra, however, and so we have further defined a subsample
of 9 very luminous and hard ULXs (see Fig.~3a): U4, U5, U10, U11, U14, U18, U19 (with
4 observations), U20, and U43. These all have luminosities in
excess of 5$\times$10$^{39}$~erg~s$^{-1}$ and photon indices
$<$1.7. This subclass is discussed further in the next sections.

\subsection{Two-Component Spectral Fits}

We next fit all spectra with the frequently used two-component model 
that consists of a MCD model plus a PL. Typical spectral states 
observed in black hole binaries and some ULXs \citep[e.g.][]{kub01} include a soft (high) state,
with a prominent blackbody component having kT~$\sim$~1~keV plus a steep
($\Gamma\sim$~2.5) PL tail, or a hard (low) state with the thermal
component being generally cooler or nonexistent and most of the
energy carried in a shallower PL ($\Gamma\sim$~1.8).
We also mention the very high state (VHS),
characterized by high luminosities, a steep PL ($\Gamma>$~2.5),
a relatively cool disk, and sometimes X-ray quasi-periodic oscillations
\citep[QPOs; see][for a detailed description]{mcc06}.

We note that these spectral states have been traditionally measured in
the 2$-$20~keV energy band and therefore may not be recognized easily
in the {\it Chandra} band. For example, in the high state the PL
component would be completely absent in our 0.3$-$8~kV band. Also, one of the most
important signatures expected from an IMBH is a cool accretion disk component.
The inner disk temperature in the MCD model scales with the black hole
mass as $\propto M^{-1/4}$. For typical values of kT$_{in}\sim$~1~keV for a black
hole binary with $10M_{\odot}$ in the high state, we would expect cool disks
with kT$_{in}\sim$~0.1$-$0.3~keV. A number of ULXs with high-quality spectra
from {\it Chandra}, {\it XMM-Newton}, and {\it RXTE} were found in the past few years
to show soft components well fit by an MCD model in this range \citep[see][]{mill04}.

To compare with published results and restrict model parameters enough to be
useful for our purposes, we select a two-component model
with fixed parameters. We assume inner disk temperatures of 0.25 or 1~keV
to represent either a cool disk or a ``normal'' disk temperature,
respectively (models PLMCD0.25 and PLMCD1.0). For ULXs, good fits are derived
for 70\% and 72\% of the sample for PLMCD0.25 and PLMCD1.0, respectively.
For the comparison sample, good fits are derived for 59\% and 55\% for
PLMCD0.25 and PLMCD1.0, respectively. The ULXs do possess a higher
percentage of good fits, but the difference is not statistically
significant given our sample sizes.
Figure~4 shows the distribution of photon indices.
The shaded areas correspond to the subsample of 9 ULXs with high luminosities
and hard X-ray spectra as defined in the previous section (see Figure~3a).
In total, there is no significant difference between the ULXs and the
lower-luminosity sources.  A K-S test for the difference 
between the distributions gives probabilities of 0.21 and 0.15 
for PLMCD0.25 and PLMCD1.0, respectively.
However, the 9 high-luminosity, hard ULXs clearly stand out.
We note the very steep PL component in some spectra for the PLMCD1.0 model.
These results correspond to the ``nonstandard model'' fits of \citet{wil06}.

We tried our two-component model with all parameters free (PLMCD model),
but many parameters are not constrained. Moreover, as seen in Figure~5,
the MCD component is very weak or practically nonexistent in many cases.
The nine high-luminosity, hard ULXs have the weakest disk components,
practically negligible.
We note a very steep PL component in some spectra here, again indicating
a nonstandard model. In these spectra the nonthermal component is soft
and strongly absorbed, as shown by the large values of the flux ratios. 
Here we only comment further on specific results for spectra that were not well fitted with
the simple models from Section 3.1. Table~\ref{table5} presents the PLMCD model 
results and in Figure~6 we plot absorbed luminosities
versus the photon index and disk temperature. The two samples
do not show significant differences. Both samples possess cool disks,
and there is no apparent correlation of the disk temperature with luminosity.
The presence of this soft disk component also causes the PL slopes
to generally become steeper, compared to our single PL fits (Figure~3).

The use of applying the F-test for an {\it added} spectral component
\citep{pro02} is controversial, so we performed simulations to check the validity of the
F-tests. For each spectrum we performed 500 simulations under a null model,
a PL in this case. We first used the command ``tclout simpars'', available in XSPEC v.12
to generate simulated parameters from the original fits.
This method uses simulations from a multivariate normal distribution
based on the covariance matrix estimated in the original fit.
The simulated F-test results are listed in parentheses in column~(7) of Table~\ref{table5}.
Any differences between the simulations and the classical F-test
are small and generally fall within the errors corresponding to the number
of simulations ($\sim$~5\%). 
The method described in  \citet{pro02} uses
a complete Bayesian Monte Carlo simulation to sample
from the posterior distribution \citep[developed by][]{van01}.
Our method {\it approximates} the posterior distribution
with a multivariate normal distribution centered at the best-fit value.
This is nevertheless better than just using the ``fakeit'' command
on the original spectrum fitted with the null model \citep[described by]
[as a ``parametric bootstrap'', and only valid when the parameters are
very well constrained]{pro02}.

In conclusion, for the subsample of 9 high-luminosity, hard ULXs (Figure~3a and
Section~3.1) we recover the same result here;  they tend to have significantly
harder spectra. We also verify that they tend to possess small contributions from
a thermal component. If such a component exists, it is practically undetectable 
with the {\it Chandra} data.
We also find that cool disks (MCD with kT$_{in}\sim$~0.1$-$0.3~keV)
are present with roughly equal probability for any given ULX
and that cool-disk components extend below the standard ULX luminosity cutoff (10$^{39}$~erg~s$^{-1}$),
down to our sample limit of 10$^{38.3}$~erg~s$^{-1}$.

\section{SHORT-TERM VARIABILITY}

Long-term flux variability from one observation to another,
which is typically years, is very common in ULXs. Short-term
flux variability, which we define here as that which can be detected
within a single observation (hours), is less frequent and is not easily
found with {\it Chandra}, probably due to limited sensitivity (i.e., not providing enough counts).
Using the K-S test, \citet{swa04} find that $\approx$15\% of
our ULXs are variable at the 95\% confidence level.

We extracted light curves for all sources, using three time bins:
3.24 (nominal frame time), 500, and 1000~s. To test for variability,
we used the K-S statistic for the nominal frame time binning, 
and the Chi-Squared test for the other two.
Using the Monte Carlo method of \citet{par06} described in the previous section,
we constructed light curves for the hardness ratios for each variable source 
and looked for variations in hardness ratios and possible time lags between
the three energy bands. We also constructed power spectra using the Leahy
normalization.

We detect variability at 95\% confidence in 6 ULXs for the longer time
frames using the Chi-Squared test, and no variability for the lower luminosity sample.
Of these, three sources were previously known to be variable. These are U2, U34, and U40.
Three other sources show variability. These are U14, U27 and C22, and the
variability scale is similar to the exposure times of the observations ($\sim$40 ks).
The K-S test identifies the same variable sources with the exception of U33 in NGC 5055,
but it finds significant variability in two additional sources:
U27 in NGC 4565 and a comparison source, C22 in NGC 6946.
There are two periodicities of 707~s detected in U33 and U6 (NGC 1313 X-1)
produced by the ACIS dither, which causes false periodic signals
at 707 and 1000~s.

We conclude that  6 ULXs are intrinsically variable, which is consistent
with the result obtained by \citet{swa04} given the small size of the sample (47 sources).
Only two sources (U2 and U40) show some energy variation, but no lag.
Given the readout time of the {\it Chandra} CCDs,
variations on timescales shorter than $\approx$10~s cannot be detected,
and features that could identify spectral states (QPOs or breaks in the power
density spectra), are not readily detectable.

\section{DISCUSSION}

From our X-ray spectral comparison between ultraluminous X-ray sources (ULXs) 
and other X-ray point sources in nearby galaxies, we find an interesting subclass of nine ULXs 
that have unique properties compared to the other sources that are classified as ULXs. 
This subclass of ULXs also differs from the lower-luminosity sample of X-ray point sources. 
While most of the ULXs we analyzed can plausibly be explained as scaled-up versions 
of Galactic black hole binaries, this particular subclass cannot. 
We discuss this subclass of ULXs followed by our general results, 
especially how our results relate to current evidence that supports the idea 
that ULXs host intermediate-mass black holes (IMBHs).

\subsection{Luminous, Hard (flat-spectrum) X-ray ULXs}

Our analysis has identified nine ULXs with very high luminosities and hard (flat) X-ray spectra. 
This sample (U4, U5, U10, U11, U14, U18, U19, U20, and U43) is shown 
in Figure~3a (upper left corner). These ULXs are all well fitted 
by a power-law (PL) model with a photon index of $<$1.7. More complex spectral fitting 
using two component models reveals that when trying to add an accretion disk component 
(the MCD model described above), the contribution of this component 
to the total X-ray flux is very small, indicating that the relative contribution 
of emission from the accretion disk to the total spectrum is small. 
This result is shown in Figure~5, where the ratio of the absorbed MCD component flux 
to the total flux is plotted against the PL index for the PLMCD model. 
By using absorbed fluxes we do not specifically show the absolute physical strength 
of the accretion disk component; however, the ratios measured in this way 
better indicate the significance of detecting a soft excess 
and are also less dependent on the modeling.

The nine members of the subclass of flat-spectrum ULXs are also very luminous, 
so they would seem to be the best candidates for hosting an IMBH 
based on a simple Eddington limit argument, which predicates 
that higher mass black holes are required to explain the most luminous accreting sources. 
The spectra of these ULXs resemble Galactic black holes in a hard state, 
but such spectral shapes are usually associated with a low-luminosity state 
in the case of Galactic black holes. If these ULXs are indeed accreting IMBHs 
in a low state (i.e. low/hard state), our result begs the question 
as to why we do not also see ULXs in a high state (high/soft state) 
with even higher luminosities. Moreover, if these are IMBHs in a low state, 
such a scenario implies very high mass black holes ($>$10$^4$~M$_{\odot}$). 
The formation of such black holes is not easy to explain.

It seems more plausible that this subclass of hard (flat) and luminous ULXs 
are accreting sources in the PL dominated very high state \citep[VHS;][]{mcc06}, 
with an unusually weak soft X-ray component. A model that describes the properties 
of a hard PL with very little flux from the disk, at least in the {\it Chandra} band, 
is the ``coupled disk-corona'' model proposed by \citet{don06}. In this model, 
the underlying accretion disk emission is distorted by a process 
that drains energy from the disk into the corona. In an extreme case, 
the inner disk emission could be almost completely Comptonized, 
and thus only the visible outer disk would contribute to the accretion disk component.  
As this obviously only appears at low temperatures it could be easily absorbed 
in some galaxies and also hard to detect. However, if our nine luminous ULXs 
are interpreted as stellar-mass black hole systems in a high state, 
we would need to explain why their X-ray spectra are much harder 
compared to those observed in our Galaxy \citep{mcc06}, 
which have typical photon indices of ~2.5 in the VHS. 
We would also need to explain how such low-mass black hole systems 
could reach such high luminosities.
A model that could explain both the flat non-thermal component 
and the weak soft disk component in luminous ULXs has been proposed by \citet{soc06}.
Their ULX model shows that at super-Eddington accretion rates, 
in the inner region of the disk, magnetic fields in the corona can prevent strong winds, 
thus the radiative efficiency is not reduced by photon trapping. 
The resulting spectrum is dominated by the coronal emission from the inner region 
and the soft thermal component is generated only in the outer disk.

There is some direct evidence that high-luminosity states in ULXs 
correlate with the hardness of the PL tail in their spectra. 
\citet{rob06} have shown this for a long {\it Chandra} observing campaign of NGC~5204 X-1 (our U36). 
These data were not available when we searched the archive. 
\citet{rob06} found that the spectrum becomes harder as the flux increases, 
over time-scales of days to weeks. The model used was a Comptonized disk model, 
and the results showed a cool disk ($\sim$0.1~keV) and an optically thick corona. 
This model demonstrates that flux variations correlate with the corona temperature. 
\citet{rob06} favor a stellar mass black hole interpretation for this ULX 
and suggest an unusual VHS, probably produced by extreme mass transfer 
from a massive star. We should note, however, that the PL slope is much steeper 
for NGC~5204 X-1 compared to the ULXs we discuss; therefore it is much easier to interpret 
as a VHS in comparison to what is observed for Galactic black holes.

It is possible that our nine ULX spectra appear to be harder than they actually are 
due to the limited energy band covered by {\it Chandra}. If these objects intrinsically possess 
a break or curvature in their spectra and the break occurs at an energy 
above the {\it Chandra} bandpass or where the sensitivity of {\it Chandra} 
falls off significantly, this might bias our modeling to measuring these sources as ``hard''. 
This appears to be the case for one ULX in our sample, NGC 1313 X-1 (U5). 
Using {\it XMM-Newton} spectra, \citet{sto06} found evidence for a break (or curvature) at 4.9~keV, 
with a photon index for the high-energy PL of 2.16 (much closer to Galactic black holes values). 
The authors show, however, that such breaks are easier to explain 
if ULXs contain stellar mass black holes rather than IMBHs. The curvature would be likely 
to originate in optically thick coronae. This theory would need to be tested 
for the remaining ULXs in our subclass by obtaining better quality spectra.

In conclusion, the subclass of nine ULXs with very high luminosities and hard (flat) X-ray spectra
suggests a PL-dominated VHS, in line with recent models of stellar mass black hole systems
in very high accretion states. The fact that the highest luminosity ULXs are explained more easily
with such models argues strongly against IMBHs as the only explanation of ULXs.

\subsection{Cool disks and the IMBH interpretation}

Using our sample of ULXs and lower luminosity X-ray sources, we have found 
that the spectral signature of a cool accretion disc is not specific to ULXs. 
The results of the widely used PLMCD model (see Fig.~6 and Table~\ref{table5}) 
show that many sources in both samples have MCD components with low inner-disk temperatures. 
Cool disks have been used until recently as support for the IMBH interpretation. 
Our results show that this evidence is nowhere near as compelling when a large sample of ULXs 
is looked at properly.

In the standard accretion disk model, cool disks are not expected for stellar-mass black holes 
accreting near their Eddington limits. The disk temperature scales with the black hole mass as
T$_{in}\propto$~M$^{-1/4}$, and is $\sim$~1~keV for stellar mass black holes. 
However, cool disks can be seen in a low (hard) state, because the temperature dependence 
on the accretion rate for standard disks is T$_{in}\propto~\dot{M}^{-1/4}$ \citep{mcc06,mill06}. 
Cool disks have indeed been found in some non-ULX sources \citep[e.g.][]{sto06}. 
The authors note the similarity of these spectral fits with those typically used for ULXs. 
They also suggest that the soft excess in some cases could be otherwise explained 
by contamination from the host galaxy.

Most sources in Figure~6b possess low disk temperatures, within both the ULX and comparison samples. 
Indeed, it is surprising that we do not see many states that are typical (high) states 
for stellar mass black hole binaries, with a prominent $\sim$1~keV blackbody component 
(although a hard PL tail would be difficult to discern in the limited {\it Chandra} bandpass). 
Only U2, U37, C3 and C9 show such high temperatures. We find a similar result 
from our cool-disk model (PLMCD0.25). A large fraction of our ULX spectra (70\%) 
are well fitted by this model, but a significant number of lower luminosity objects (59\%) are as well.

There are theoretical models that do not require the presence of an IMBH 
to explain a cool disk at high accretion rates and high luminosities (i.e., for ULXs). 
We already mentioned the model proposed by \citet{don06}, that explains cool disks 
by a process of draining energy from the disk to launch an optically-thick corona 
that obscures the hot inner regions of the disk. \citet{fre06} developed microblazar models 
with magnetized jets that cause a transference of disk energy into the jet, 
thus making the disk fainter and cooling it at the same time. 
Other phenomenological models include the ``dual thermal'' model of \citet{wil06}, 
in which the soft excess comes from an optically thick outflow 
produced at high accretion rates \citep[see][]{king03}, which is seen 
in addition to a disk component with a temperature similar to those seen 
in stellar-mass black hole binaries. 
This latter model was proposed to explain the alternate model of \citet{wil06}.

\citet{don06} found evidence of a Galactic black hole that supports the interpretation of ULXs 
as stellar mass black holes in a VHS. The microquasar XTE J1550-564 has a ``strong VHS'' 
\citep[see also][]{kub04}, where the disc temperature decreases with luminosity, 
reaching values of 0.3$-$0.4~keV. 
A similar behavior has been found for NGC~1313 X-2 \citep{fen07}.
This suggests a new type of VHS, 
a so-called ultraluminous branch, which is very similar to the ULX spectra
\citep{rob07,sor07}. In this interpretation, ULXs represent the high end 
of such an accretion state, with black hole masses up to 100~M$_{\odot}$ and accretion rates 
up to 20 times the Eddington limit. Forming black holes with such masses 
is much easier to explain than forming IMBHs. For example \citet{bel04} 
showed that black holes with masses of 80~M$_{\odot}$ or more can form through binary mergers. 
\citet{sorw06} suggested that black holes of up to 200~M$_{\odot}$ could form 
by large-scale dynamical collapse of protoclusters in active regions in galaxies. 
These formation mechanisms are supported by the association between ULXs, 
star-forming regions and colliding galaxies.

\subsection{Conclusion}

We have found that the highest luminosity ULXs tend to have the hardest X-ray spectra 
in the {\it Chandra} bandpass and are well fitted by a simple power-law model, 
without evidence for thermal accretion disc components.  Such spectra 
are not consistent with current IMBH models, but are more in line with current models 
of extreme very high states, or perhaps a new ``ultraluminous state'' \citep{rob07}, 
in stellar mass black holes.

Our work shows that cool accretion disks are not exclusive to the ULX 
class, suggesting that low-temperature IMBHs are 
not the only explanation for this phenomenon.  In general, our results 
show that ULXs are likely to be composed of several distinct types of 
objects, and that these types may extend into lower X-ray luminosity 
classes, such as classical Galactic black hole candidates and other 
objects in our comparison sample.  Our conclusions provide another ``nail 
in the coffin'' for assumptions that ULXs are simply a class of accreting 
IMBHs.

No other specific properties have been found for the ULX group,
except for spectral hardening at the highest luminosities. 
All these results suggest that ULXs are the highest luminosity end of
stellar mass black hole binaries, with the largest black holes permitted by current 
formation mechanisms and/or accreting at super-Eddington rates.

\acknowledgments

This research has made
use of the SAOImage DS9, developed by Smithsonian Astrophysical
Observatory and of the NASA/IPAC Extragalactic Database (NED)
which is operated by the Jet Propulsion Laboratory, California
Institute of Technology, under contract with the National Aeronautics
and Space Administration.


\clearpage
\pagestyle{empty} 
\begin{deluxetable}{r|cccccrrccccll}
\tablecolumns{13}                                    
\rotate                                     
\tablewidth{0pt} 
\tabletypesize{\tiny} 
\tablenum{1}              
\tablecaption{Piled-up observations\label{table1}}          
\tablehead{
\colhead{Obs.} & \colhead{Name} & \colhead{Position} & \colhead{Galaxy} &
\colhead{Log L$_X$} & \colhead{OBSID} & \colhead{Date} &  
\colhead{CCD} & \colhead{$\theta_{off}$} & \colhead{Count rate} & \colhead{Reduced rate} & 
\colhead{Pile-up} & \colhead{Alternate names} & \colhead{Ref} \\ 
\colhead{(1)} & \colhead{(2)} & \colhead{(3)} & \colhead{(4)} & \colhead{(5)} & 
\colhead{(6)} & \colhead{(7)} & 
\colhead{(8)} & \colhead{(9)} & \colhead{(10)} & \colhead{(11)} & \colhead{(12)} 
& \colhead{(13)} & \colhead{(14)} \\
}
\startdata
\multicolumn{14}{c}{ULXs}\\
\tableline

1 & U5  & X031820.0$-$662911 & NGC 1313    & 39.47 & 2950 & 2002 Oct 13 & 7(1/1) & 2.4 & 0.25 & 0.13 & 0.15 &                   &               \\
2 & -   & X081929.0$+$704219 & Holmberg II & 39.91 & 1564 & 2001 Nov 02 & 7(1/4) & 0.6 & 0.51 & 0.18 & 0.21 & IXO 31, ULX1, X-1 & 1, 2, 3       \\
3 & U10 & X095550.0$+$694046 & M82         & 39.52 & 1302 & 1999 Sep 20 & 3(1/1) & 0.4 & 0.12 & 0.12 & 0.14 &                   &               \\
4 & U10 & X095550.0$+$694046 &             & 39.51 & 361  & 1999 Sep 20 & 3(1/1) & 0.4 & 0.11 & 0.11 & 0.13 &                   &               \\
5 & U10 & X095550.0$+$694046 &             & 40.11 & 379  & 2000 Mar 11 & 3(1/1) & 4.2 & 0.46 & 0.09 & 0.11 &                   &               \\
6 & U36 & X132938.6$+$582505 & NGC 5204    & 39.85 & 2028 & 2001 Jan 09 & 7(1/8) & 0.6 & 0.43 & 0.10 & 0.12 &                   &               \\
7 & -   & X140319.6$-$412258 & NGC 5408    & 39.82 & 2885 & 2002 May 07 & 7(1/4) & 0.7 & 0.32 & 0.08 & 0.11 & NGC 5408 X-1      & 4, 5, 6       \\
8 & U43 & X141312.2$-$652014 & Circinus    & 39.30 & 356  & 2000 Mar 14 & 7(1/1) & 0.4 & 0.12 & 0.12 & 0.15 &                   &               \\

\tableline                                                                                                 
\multicolumn{14}{c}{Comparison}\\  
\tableline

9 & -   & X095533.0$+$690033 & M81         & 38.39 & 735  & 2000 May 07 & 7(1/1) & 1.0 & 0.18 & 0.59 & 0.21 & MF97              & 1, 4, 5, 6, 7 \\

\enddata

\tablecomments{                                                    
(1) Observation number;                                                 
(2) Sample source name if the same as one in Table 2;
(3) X-ray positions (J2000);
(4) Host galaxy;
(5) Approximate observed luminosity in units of erg~s$^{-1}$, in the energy band 0.3$-$8.0~keV derived 
from the count rate of the reprocessed data; we assumed 
a PL model with $\Gamma=$~1.8 and Galactic absorption column;
(6) Observation ID;
(7) Date of observation start;
(8) CCD number where the object is located and subarray values (in parantheses);
(9) Off axis angle of the source in arcminutes;
(10) Count rate in s$^{-1}$;
(11) Reduced count rate calculated for pileup estimations explained in Section~2.4.
This takes into account the off-axis angle in col. (9) and the subarray values in col. (8);
(12) Pile-up estimation based on the reduced count rate in column 11;
(13) Common names from the literature in col. (14) (see Table~2 for common names and references for objects listed in col. 2);
(14) References.
}
\tablerefs{
1. \citet{liub05};
2. \citet{col02}; 
3. \citet{goad06};
4. \citet{fen05};
5. \citet{lium05};
6. \citet{swa04};
7. \citet{swa03};
}
\end{deluxetable}


\clearpage
\pagestyle{empty}

\begin{deluxetable}{rcccccrrccccclcl}
\tablecolumns{15}                                    
\rotate    
\tablewidth{0pt} 
\tabletypesize{\tiny}
\setlength{\tabcolsep}{0.05in}
\tablenum{2} 
\tablecaption{Properties of sample objects\label{table2}}          
\tablehead{
\colhead{No} & \colhead{Name} & \colhead{Position} & \colhead{Galaxy} & 
\colhead{Dist} & \colhead{N$^{Gal}_H$} & \colhead{OBS} & 
\colhead{CCD} & \colhead{Log L$_X$} & \colhead{Date} & \colhead{Exp} & 
\colhead{Counts} & \colhead{$\theta_{off}$} & 
\colhead{Alternate names} & \colhead{Loc} & \colhead{Ref} \\ 
\colhead{(1)} & \colhead{(2)} & \colhead{(3)} & \colhead{(4)} & \colhead{(5)} & 
\colhead{(6)} & \colhead{(7)} & 
\colhead{(8)} & \colhead{(9)} & \colhead{(10)} & \colhead{(11)} & \colhead{(12)} & 
\colhead{(13)} & \colhead{(14)} & \colhead{(15)} & \colhead{(16)}
}
\startdata
\multicolumn{14}{c}{\small {ULX sample}}\\
\tableline

1  & U1   & X012435.2$+$034731 & NGC 520    & 29.6 & 3.3    & 2924 & 7(1/1) & 40.1 & 2003 Jan 29 & 49.3 & 1037.2   $\pm$ 32.3  & 2.1  & Source 11 	               & DB  & 1			  \\
2  & U2   & X013350.9$+$303938 & M33        & 0.7  & 5.69   & 787  & 7(1/4) & 39.0 & 2000 Jan 11 & 9.3  & 26733.4  $\pm$ 163.6 & 8.9  & M33 X-8	                       & N   & 2, 3, 4, 5		  \\
3  &      & X013350.9$+$303939 &            &      & 5.69   & 2023 & 7(1/1) & 38.9 & 2001 Jul 06 & 88.8 & 171301.0 $\pm$ 415.2 & 12.5 &			               &     &  			  \\
5  & U3   & X022231.4$+$422024 & NGC 891    & 9.6  & 8.12   & 794  & 7(1/1) & 39.4 & 2000 Nov 01 & 50.9 & 1977.4   $\pm$ 44.5  & 1.7  & NGC 891 X-4 	               & DB  & 6, 7			  \\
6  & U4   & X024238.9$-$000055 & M77        & 15.2 & 3.54   & 344  & 7(1/1) & 39.7 & 2000 Feb 21 & 47.4 & 1524.8   $\pm$ 39.6  & 0.7  & -		               & A   & 8			  \\
7  & U5   & X031820.0$-$662911 & NGC 1313   & 3.7  & 3.96   & 3550 & 2(1/1) & 40.1 & 2002 Nov 09 & 14.6 & 10486.7  $\pm$ 102.7 & 6.0  & IXO 7, XMM1, NGC 1313 X-1      & B   & 9, 64, 10, 11		  \\
8  & U6   & X034555.7$+$680455 & IC 342     & 3.9  & 29.39  & 2916 & 7(1/8) & 39.5 & 2002 Apr 29 & 9.3  & 2033.6   $\pm$ 45.1  & 0.5  & IXO 22, IC 342 X-7, XMM1, X-1  & A   & 9, 12, 64, 6, 13, 26, 47   \\
9  &      & X034555.6$+$680456 &            &      & 29.39  & 2917 & 7(1/8) & 39.5 & 2002 Aug 26 & 9.9  & 2191.8   $\pm$ 46.8  & 0.6  &				       &     &  			  \\
10 & U7   & X073625.5$+$653540 & NGC 2403   & 4.2  & 4.17   & 2014 & 7(1/1) & 39.2 & 2001 Apr 17 & 35.6 & 5364.2   $\pm$ 73.3  & 2.7  & Source 21, NGC 2403 X-1, XMM1  & A   & 14, 6, 64		  \\
11 & U8   & X085333.7$+$511930 & NGC 2681   & 13.3 & 2.48   & 2061 & 7(1/1) & 39.2 & 2001 May 02 & 79.0 & 1105.9   $\pm$ 33.3  & 1.3  & NGC 2681 PSX-3	   	       & D   & 15			  \\
12 & U9   & X095546.5$+$694040 & M82        & 5.2  & 4.02   & 361  & 3(1/1) & 39.0 & 1999 Sep 20 & 33.3 & 1174.2   $\pm$ 34.9  & 0.8  & Source 9	               & SF  & 16			  \\
13 & U10  & X095550.1$+$694048 &            &      & 4.03   & 378  & 3(1/1) & 40.0 & 1999 Dec 30 & 4.1  & 1404.7   $\pm$ 38.0  & 4.0  & Source 7, M82 X-1	       & SF  & 16, 17, 29		  \\
14 & U11  & X095551.0$+$694045 &            &      & 4.03   & 2933 & 7(1/1) & 39.2 & 2002 Jun 18 & 18.0 & 1595.2   $\pm$ 41.3  & 0.6  & Source 5	               & SF  & 16			  \\
15 & U12  & X095551.1$+$694043 &            &      & 4.03   & 361  & 3(1/1) & 39.1 & 1999 Sep 20 & 33.3 & 1353.6   $\pm$ 39.1  & 0.4  & Source 4	               & SF  & 16			  \\
16 & U13  & X103843.3$+$533102 & NGC 3310   & 18.7 & 1.12   & 2939 & 7(1/2) & 39.7 & 2003 Jan 25 & 47.2 & 1003.9   $\pm$ 31.7  & 0.3  & IXO 38, NGC 3310 ULX2, X-3     & A   & 9, 18, 19		  \\
17 & U14  & X103845.9$+$533012 &            &      & 1.11   & 2939 & 7(1/2) & 39.8 & 2003 Jan 25 & 47.2 & 1541.8   $\pm$ 41.0  & 0.6  & NGC 3310 X-1, X1	       & N   & 6, 18			  \\
18 & U15  & X103846.0$+$533004 &            &      & 1.11   & 2939 & 7(1/2) & 39.8 & 2003 Jan 25 & 47.2 & 1221.6   $\pm$ 35.7  & 0.7  & -			       & SF  & -			  \\
19 & U16  & X111126.0$+$554017 & M108       & 14.1 & 0.78   & 2025 & 7(1/1) & 39.4 & 2001 Sep 08 & 59.4 & 1278.9   $\pm$ 35.9  & 2.8  & Source 26		       & D   & 20			  \\
20 & U17  & X112015.8$+$133514 & NGC 3628   & 7.7  & 2.22   & 2039 & 7(1/1) & 39.3 & 2000 Dec 02 & 58.0 & 2995.8   $\pm$ 54.8  & 0.9  & IXO 39		               & DB  & 9, 21			  \\
21 & U18  & X120151.4$-$185225 & NGC 4038/9 & 21.7 & 3.95   & 3040 & 7(1/1) & 39.7 & 2001 Dec 29 & 69.0 & 1009.7   $\pm$ 31.9  & 0.9  & Source 11		       & AM  & 22, 23			  \\
22 &      & X120151.3$-$185225 &            &      & 3.95   & 3043 & 7(1/1) & 39.8 & 2002 Apr 18 & 67.1 & 1377.4   $\pm$ 37.3  & 1.0  &				       &     &  			  \\
23 &      & X120151.3$-$185225 &            &      & 3.95   & 3041 & 7(1/1) & 39.8 & 2002 Nov 22 & 72.9 & 1491.4   $\pm$ 38.8  & 0.9  &				       &     &  			  \\
24 & U19  & X120152.1$-$185134 &            &      & 3.95   & 315  & 7(1/1) & 39.9 & 1999 Dec 01 & 72.2 & 1984.1   $\pm$ 44.6  & 1.6  & Source 16		       & AM  & 22, 23			  \\
25 &      & X120152.1$-$185133 &            &      & 3.95   & 3040 & 7(1/1) & 39.8 & 2001 Dec 29 & 69.0 & 1587.0   $\pm$ 40.0  & 0.9  & 		               &     &  			  \\
26 &      & X120152.1$-$185133 &            &      & 3.95   & 3042 & 7(1/1) & 39.8 & 2002 May 31 & 67.3 & 1474.9   $\pm$ 38.5  & 1.6  &				       &     &  			  \\
27 &      & X120152.1$-$185133 &            &      & 3.95   & 3041 & 7(1/1) & 39.8 & 2002 Nov 22 & 72.9 & 1491.8   $\pm$ 38.8  & 0.8  &				       &     &  			  \\
28 & U20  & X120155.6$-$185215 &            &      & 3.96   & 315  & 7(1/1) & 39.8 & 1999 Dec 01 & 72.2 & 1344.0   $\pm$ 37.3  & 1.9  & Source 42		       & AM  & 22, 23			  \\
29 & U21  & X120156.4$-$185158 &            &      & 3.96   & 315  & 7(1/1) & 39.7 & 1999 Dec 01 & 72.2 & 1307.1   $\pm$ 36.2  & 1.6  & Source 44		       & AM  & 22, 23			  \\
30 &      & X120156.5$-$185157 &            &      & 3.96   & 3040 & 7(1/1) & 39.7 & 2001 Dec 29 & 69.0 & 1264.8   $\pm$ 35.6  & 0.4  &				       &     &  			  \\
31 & U22  & X120807.5$+$651028 & NGC 4125   & 18.1 & 1.82   & 2071 & 7(1/1) & 39.5 & 2001 Sep 09 & 64.2 & 1051.7   $\pm$ 33.0  & 0.4  & -			       & E   & 68			  \\
32 & U23  & X123030.6$+$414142 & NGC 4485   & 9.3  & 1.78   & 1579 & 7(1/1) & 39.6 & 2000 Nov 03 & 19.5 & 1450.1   $\pm$ 38.1  & 2.6  & IXO 62, NGC 4485 X-1           & A   & 9, 6, 24 		  \\
33 & U24  & X123049.2$+$122604 & M87        & 17.1 & 2.54   & 2707 & 7(1/1) & 39.3 & 2002 Jul 06 & 98.7 & 1064.6   $\pm$ 36.8  & 3.1  & -			       & E   & 68			  \\
34 & U25  & X123551.7$+$275604 & NGC 4559   & 9.7  & 0.82   & 2026 & 7(1/4) & 39.9 & 2001 Jan 14 & 9.4  & 1434.4   $\pm$ 37.9  & 0.6  & IXO 65, NGC 4559 X-1, X7       & D   & 9, 6, 25, 26, 48 	  \\
35 &      & X123551.7$+$275604 &            &      & 0.82   & 2027 & 7(1/4) & 40.1 & 2001 Jun 04 & 10.7 & 2093.2   $\pm$ 45.8  & 0.6  &				       &     &  			  \\
36 & U26  & X123558.6$+$275742 &            &      & 0.8    & 2027 & 7(1/4) & 39.8 & 2001 Jun 04 & 10.7 & 1300.9   $\pm$ 36.1  & 2.9  & IXO 66, NGC 4559 X-4, X10      & DB  & 9, 6, 25, 26, 48 	  \\
37 & U27  & X123617.4$+$255856 & NGC 4565   & 16.4 & 1.31   & 3950 & 7(1/1) & 39.8 & 2003 Feb 08 & 59.2 & 2146.5   $\pm$ 46.5  & 2.0  & IXO 67, NGC 4565 ULX4          &  B  & 9, 27, 28		  \\
38 & U28  & X123740.3$+$114728 & NGC 4579   & 20.3 & 2.52   & 807  & 7(1/4) & 40.1 & 2000 May 02 & 33.9 & 1654.6   $\pm$ 40.7  & 1.3  & NGC 4579 X-1                   & D   & 30			  \\
39 & U29  & X124155.6$+$323217 & NGC 4631   & 6.9  & 1.29   & 797  & 7(1/1) & 39.2 & 2000 Apr 16 & 59.2 & 3223.1   $\pm$ 56.8  & 0.5  & IXO 68, NGC 4631 X-1, XMM1     & SF  & 9, 6, 64 		  \\
40 & U30  & X125053.3$+$410714 & M94        & 4.3  & 1.44   & 808  & 7(1/4) & 39.0 & 2000 May 13 & 47.4 & 4472.6   $\pm$ 70.8  & 0.7  & NGC 4736 X-1		       & DB  & 30			  \\
41 & U31  & X130521.9$-$492827 & NGC 4945   & 5.2  & 14.94  & 864  & 7(1/1) & 39.1 & 2000 Jan 27 & 49.1 & 2983.5   $\pm$ 54.9  & 1.3  & NGC 4945 XMM4		       & DB  & 64			  \\
42 & U32  & X130532.9$-$492734 &            &      & 14.84  & 864  & 7(1/1) & 39.1 & 2000 Jan 27 & 49.1 & 2797.7   $\pm$ 53.2  & 0.7  & NGC 4945 X-2, XMM1             & DB  & 31, 64			  \\
43 & U33  & X131519.5$+$420302 & NGC 5055   & 7.2  & 1.3    & 2197 & 7(1/1) & 39.8 & 2001 Aug 27 & 28.0 & 2354.6   $\pm$ 48.6  & 6.0  & IXO 74, NGC 5055 X-2           & D   & 9, 6			  \\
44 & U34  & X132507.4$-$430410 & Cen A      & 4.9  & 8.41   & 316  & 3(1/1) & 39.0 & 1999 Dec 05 & 35.7 & 1108.8   $\pm$ 33.9  & 9.2  & IXO 75			       & E   & 9, 32			  \\
45 &      & X132507.5$-$430410 &            &      & 8.41   & 962  & 1(1/1) & 39.3 & 2000 May 17 & 36.5 & 2556.1   $\pm$ 50.7  & 5.5  &				       &     &  			  \\
46 & U35  & X132519.8$-$430317 &            &      & 8.4    & 316  & 3(1/1) & 39.2 & 1999 Dec 05 & 35.7 & 2124.0   $\pm$ 46.4  & 7.1  & IXO 76			       & E   & 9, 32, 63		  \\
47 & U36  & X132938.6$+$582506 & NGC 5204   & 4.8  & 1.38   & 2029 & 7(1/8) & 39.4 & 2001 May 02 & 9.0  & 1498.1   $\pm$ 38.7  & 0.6  & IXO 77, NGC 5204 X-1, XMM1     & SF  & 9, (6, 18, 26, 33, 34), 64 \\
48 & U37  & X133719.8$-$295349 & M83        & 4.7  & 3.69   & 793  & 6(1/1) & 39.0 & 2000 Apr 29 & 51.0 & 2419.2   $\pm$ 49.2  & 2.7  & IXO 82, H30, XMM1	       & D   & 9,  35, 64		  \\
49 & U38  & X140304.0$+$542735 & M101       & 5.4  & 1.15   & 4731 & 6(1/1) & 39.2 & 2004 Jan 19 & 56.2 & 3213.5   $\pm$ 56.8  & 4.4  & MF37, ULX2, H19, XMM-1, XMM2   & A   & 39, 18, 37, 40, 64	  \\
50 & U39  & X140314.3$+$541806 &            &      & 1.15   & 5309 & 7(1/1) & 39.0 & 2004 Mar 14 & 70.8 & 3889.1   $\pm$ 62.6  & 5.2  & H25, P51, XMM-2, XMM1          & A   & 37, 38, 40, 64		  \\
51 &      & X140314.3$+$541806 &            &      & 1.15   & 4732 & 7(1/1) & 39.0 & 2004 Mar 19 & 69.8 & 3902.9   $\pm$ 62.8  & 5.2  &				       &     &  			  \\
52 & U40  & X140332.4$+$542103 &            &      & 1.15   & 934  & 7(1/1) & 39.2 & 2000 Mar 26 & 98.2 & 9024.5   $\pm$ 95.1  & 3.8  & M101 X5, H32, P98, ULX-1       & A   & 18, 37, 38, 49, 65, 66, 60 \\
53 & U41  & X140414.3$+$542604 &            &      & 1.15   & 934  & 3(1/1) & 39.1 & 2000 Mar 26 & 98.2 & 3549.6   $\pm$ 59.7  & 10.6 & IXO 83, ULX3, H45, XMM-3       & D   & 9, 18, 37, 40		  \\
54 &      & X140414.1$+$542603 &            &      & 1.15   & 4731 & 2(1/1) & 38.8 & 2004 Jan 19 & 56.2 & 1085.5   $\pm$ 33.1  & 8.7  &				       &     &  			  \\
55 &      & X140414.2$+$542603 &            &      & 1.15   & 5300 & 3(1/1) & 39.2 & 2004 Mar 07 & 52.1 & 2306.5   $\pm$ 48.2  & 11.0 &				       &     &  			  \\
56 &      & X140414.2$+$542603 &            &      & 1.15   & 5309 & 3(1/1) & 38.9 & 2004 Mar 14 & 70.8 & 1592.2   $\pm$ 40.2  & 11.5 &				       &     &  			  \\
57 &      & X140414.2$+$542603 &            &      & 1.15   & 4732 & 3(1/1) & 38.8 & 2004 Mar 19 & 69.8 & 1147.7   $\pm$ 34.3  & 11.5 &				       &     &  			  \\
58 & U42  & X141310.1$-$652045 & Circinus   & 3.7  & 59.7   & 356  & 7(1/1) & 39.1 & 2000 Mar 14 & 23.1 & 1715.3   $\pm$ 41.4  & 0.9  & CG X-2, source F	       & D   & 41, 42			  \\
59 & U43  & X141312.2$-$652014 &            &      & 59.92  & 365  & 7(1/8) & 39.7 & 2000 Mar 14 & 5.0  & 1634.5   $\pm$ 40.5  & 0.4  & CG X-1, source J	       & D   & 41, 42, 43		  \\
60 & U44  & X145358.9$+$033217 & NGC 5775   & 22.4 & 3.51   & 2940 & 7(1/1) & 39.9 & 2002 Apr 05 & 58.2 & 1324.2   $\pm$ 36.4  & 1.1  & -			       & D   &  			  \\
61 & U45  & X203500.7$+$601131 & NGC 6946   & 5.5  & 20.23  & 1043 & 7(1/1) & 39.6 & 2001 Sep 07 & 58.3 & 8451.9   $\pm$ 92.1  & 4.8  & MF16, NGC 6946 X-11, 58, X8    & D   & 44, 6, 45, 46		  \\
62 &      & X203500.8$+$601131 &            &      & 20.23  & 4404 & 7(1/1) & 39.5 & 2002 Nov 25 & 30.0 & 3750.0   $\pm$ 61.3  & 2.9  &				       &     &  			  \\
63 & U46  & X225724.7$-$410344 & NGC 7424   & 11.5 & 1.33   & 3496 & 7(1/1) & 39.7 & 2002 Jun 11 & 23.9 & 1370.8   $\pm$ 37.0  & 2.2  & ULX2			       & A   & 67			  \\
64 & U47  & X225728.9$-$410212 &            &      & 1.32   & 3496 & 7(1/1) & 39.7 & 2002 Jun 11 & 23.9 & 1331.9   $\pm$ 36.5  & 0.5  & ULX1			       & D   & 67			  \\

\tableline                              
\multicolumn{14}{c}{\small {Comparison sample}}\\ 
\tableline

1  & C1   & X001528.9$-$391319 & NGC 55     & 1.3  & 1.74  & 2255 & 0(1/1) & 38.43 & 2001 Sep 11 & 59.4 & 8553.3   $\pm$ 92.5  & 3.7  & Source 7, 6, N55	      & D   & 50, 7, 51, 52		 \\
2  & C2   & X004238.5$+$411604 & M31        & 0.7  & 6.66  & 1585 & 0(1/1) & 38.30 & 2001 Nov 19 & 4.9  & 1806.5   $\pm$ 42.6  & 4.3  & r2-26, source 35	      & DB  & (53, 62), (54, 55, 56)	 \\
3  &      & X004238.5$+$411604 &            &      & 6.66  & 2895 & 0(1/1) & 38.39 & 2001 Dec 07 & 4.9  & 2130.6   $\pm$ 46.3  & 5.3  &				      &     &				 \\
4  &      & X004238.6$+$411603 &            &      & 6.66  & 2896 & 1(1/1) & 38.40 & 2002 Feb 06 & 4.9  & 2302.4   $\pm$ 48.1  & 5.2  &				      &     &				 \\
5  &      & X004238.6$+$411604 &            &      & 6.66  & 2898 & 3(1/1) & 38.48 & 2002 Jun 02 & 4.9  & 2384.3   $\pm$ 49.0  & 6.8  &				      &     &				 \\
6  & C3   & X004305.7$+$411703 &            &      & 6.74  & 1575 & 7(1/1) & 38.27 & 2001 Oct 05 & 37.7 & 20560.8  $\pm$ 143.5 & 4.8  & -			      & DB  & 62, 55			 \\
7  & C4   & X004722.6$-$252051 & NGC 253    & 3.0  & 1.35  & 790  & 6(1/1) & 38.36 & 1999 Dec 27 & 43.5 & 1022.2   $\pm$ 32.3  & 7.8  & NGC 253 PSX-5, X21, XMM2      & A   & 15, (57, 58, 61), 64	 \\
8  & C5   & X004733.0$-$251749 &            &      & 1.37  & 969  & 7(1/1) & 38.67 & 1999 Dec 16 & 14.0 & 1150.2   $\pm$ 34.0  & 0.3  & NGC 253 PSX-2, X33, XMM1      &     & 15, (57, 58), 64  	 \\
9  &      & X004733.0$-$251749 &            &      & 1.37  & 790  & 6(1/1) & 38.85 & 1999 Dec 27 & 43.5 & 3246.7   $\pm$ 57.4  & 5.5  &				      &     &				 \\
10 & C6   & X004735.2$-$251512 &            &      & 1.39  & 790  & 6(1/1) & 38.57 & 1999 Dec 27 & 43.5 & 1811.8   $\pm$ 42.6  & 3.8  & NGC253  PSX-7, X36, XMM3      & A   & 15, (57, 58), 64  	 \\
11 & C7   & X073655.6$+$653541 & NGC 2403   & 4.2  & 4.17  & 2014 & 7(1/1) & 38.94 & 2001 Apr 17 & 35.6 & 2608.9   $\pm$ 51.1  & 1.1  & Source 20, XMM3		      & DB  & 14, 64			 \\
12 & C8   & X073702.4$+$653935 &            &      & 4.18  & 2014 & 7(1/1) & 38.72 & 2001 Apr 17 & 35.6 & 1600.1   $\pm$ 40.1  & 5.0  & Source 1, NGC 2403 X-4, XMM4  & A   & 14, 6, 64 		 \\
13 & C9   & X122809.3$+$440508 & NGC 4449   & 3.0  & 1.5   & 2031 & 7(1/1) & 38.38 & 2001 Feb 04 & 26.6 & 1138.8   $\pm$ 33.8  & 2.3  & NGC 4449 X-1, source 10       & SF  & 6, 59			 \\
14 & C10  & X122817.8$+$440634 &            &      & 1.49  & 2031 & 7(1/1) & 38.46 & 2001 Feb 04 & 26.6 & 1356.9   $\pm$ 36.9  & 1.7  & NGC 4449 X-7, source 27	      & SF  & 6, 59			 \\
15 & C11  & X124211.1$+$323236 & NGC 4631   & 6.9  & 1.29  & 797  & 7(1/1) & 38.73 & 2000 Apr 16 & 59.2 & 1104.0   $\pm$ 33.3  & 3.1  & NGC 4631 PSX-1, XMM5	      & D   & 15, 64			 \\
16 & C12  & X125050.3$+$410712 & M94        & 4.3  & 1.44  & 808  & 7(1/4) & 38.51 & 2000 May 13 & 47.4 & 1349.6   $\pm$ 37.0  & 0.3  & M94 X-4			      & DB  & 30			 \\
17 & C13  & X125052.7$+$410719 &            &      & 1.44  & 808  & 7(1/4) & 38.79 & 2000 May 13 & 47.4 & 2553.3   $\pm$ 51.6  & 0.6  & M94 X-3			      & DB  & 30			 \\
18 & C14  & X125053.1$+$410712 &            &      & 1.44  & 808  & 7(1/4) & 38.83 & 2000 May 13 & 47.4 & 2782.9   $\pm$ 53.6  & 0.6  & M94 X-2			      & DB  & 30			 \\
19 & C15  & X130518.5$-$492824 & NGC 4945   & 5.2  & 14.96 & 864  & 7(1/1) & 38.69 & 2000 Jan 27 & 49.1 & 1115.3   $\pm$ 33.8  & 1.8  & NGC 4945 XMM3		      & DB  & 64			 \\
20 & C16  & X130538.1$-$492545 &            &      & 14.74 & 864  & 6(1/1) & 38.94 & 2000 Jan 27 & 49.1 & 1535.9   $\pm$ 39.3  & 2.6  & Source 3, NGC 4945 XMM2	      & D   & 31, 64			 \\
21 & C17  & X133659.5$-$294959 & M83        & 4.7  & 3.69  & 793  & 7(1/1) & 38.54 & 2000 Apr 29 & 51.0 & 1249.9   $\pm$ 35.5  & 3.9  & H17, source 28, M83 XMM2      & A   & 35, 36, 64		 \\
22 & C18  & X133700.9$-$295203 &            &      & 3.7   & 793  & 7(1/1) & 38.58 & 2000 Apr 29 & 51.0 & 1341.9   $\pm$ 40.8  & 2.0  & source 44		      & DB  & 36			 \\
23 & C19  & X133704.3$-$295404 &            &      & 3.72  & 793  & 7(1/1) & 38.59 & 2000 Apr 29 & 51.0 & 1381.1   $\pm$ 37.2  & 1.0  & H26, source 62		      & A   & 35, 36			 \\
24 & C20  & X133704.4$-$295122 &            &      & 3.69  & 793  & 7(1/1) & 38.63 & 2000 Apr 29 & 51.0 & 1527.7   $\pm$ 39.2  & 2.2  & H27, source 64, M83 XMM3      & DB  & 35, 36, 64		 \\
25 & C21  & X140228.3$+$541627 & M101       & 5.4  & 1.14  & 5322 & 6(1/1) & 38.76 & 2004 May 03 & 64.7 & 1235.2   $\pm$ 35.3  & 5.6  & M101 XMM4		      & D   & 64			 \\
26 & C22  & X203500.1$+$600908 & NGC 6946   & 5.5  & 20.13 & 1043 & 7(1/1) & 38.91 & 2001 Sep 07 & 58.3 & 1894.7   $\pm$ 43.6  & 3.4  & IXO 85, NGC 6946 X-9, 56, X7  & A   & 9, 6, 45, 46		 \\
\enddata

\tablecomments{                                                    
(1) Observation number;                                                 
(2) Source name;
(3) X-ray positions (J2000);
(4) Host galaxy;
(5) Galaxy distance from Tully(1988) in Mpc;
(6) Galactic absorption column in units of 10$^{20}$cm$^{-2}$;
(7) Observation ID;
(8) CCD number where the object is located and subarray values (in parantheses); 
the subarray value represents the fraction of the CCD actually used in the observation;
(9) Approximate observed luminosity in units of erg s$^{-1}$, in the energy band 0.3$-$8.0~keV derived 
from the count rate of the reprocessed data; we assumed 
a PL model with $\Gamma=$~1.8 and Galactic absorption column;
(10) Date of observation start;
(11) Exposure time in ks;
(12) Net counts in the 0.3$-$8.0 keV energy band;
(13) Off axis angle of the source in arcminutes; the values listed here 
and the subarray values in col. (8) were used when we rejected the piledup sources;
(14) Common names from the literature in column 16 (the names correspond to 
references in the same order; references that use the same name are in parantheses; some papers do not give special names or the names 
are given using the coordinates, these were not used);
(15) Location in the galaxy; abbreviations are: A - spiral arm, D - disk, DB - disk or bulge, E - elliptical galaxy, no special location, 
SF - star forming region, AM - arm in merger, N -  nucleus;
(16) References
}
\tablerefs{
1. \citet{read05};
2. \citet{dub02};
3. \citet{col99};
4. \citet{fos04};
5. \citet{lapa03};
6. \citet{rob2000};
7. \citet{read97};
8. \citet{smi03};
9. \citet{col02};
10. \citet{col95};
11. \citet{mill04};
12. \citet{kong03};
13. \citet{sug01};
14. \citet{sch03};
15. \citet{hum03};
16. \citet{mat01};
17. \citet{stro03};
18. \citet{liub05};
19. \citet{jen04a};
20. \citet{wang03};
21. \citet{str01};
22. \citet{fab01};
23. \citet{zez02};
24. \citet{rob02};
25. \citet{vog97};
26. \citet{rob04};
27. \citet{fos04};
28. \citet{wu02}; 
29. \citet{muc06};
30. \citet{era02};
31. \citet{gua00};
32. \citet{kra01};
33. \citet{rob01};
34. \citet{rob05};
35. \citet{imm99};
36. \citet{sor03};
37. \citet{wang99};
38. \citet{pen01};
39. \citet{mat97};
40. \citet{jen04b};
41. \citet{bau01};
42. \citet{smi01};
43. \citet{wei04};
44. \citet{roco03};
45. \citet{hol03};
46. \citet{lira00};
47. \citet{rob03};
48. \citet{cro04}; 
49. \citet{muk03};
50. \citet{sch97};
51. \citet{rob97};
52. \citet{sto04};
53. \citet{kong02};
54. \citet{pri93};
55. \citet{kaa02};
56. \citet{bar03};
57. \citet{vog99};
58. \citet{pie01};
59. \citet{sum03};
60. \citet{kun05};
61. \citet{tan05};
62. \citet{wil04};
63. \citet{gho06};
64. \citet{win06};
65. \citet{muk05};
66. \citet{kong05};
67. \citet{sor06};
68. \citet{swa04};
}
\end{deluxetable}


\begin{deluxetable}{l|ccccc|ccccc}
\tablecolumns{10} 
\tabletypesize{\scriptsize}                                                                 
\tablewidth{0pt}                                                            
\setlength{\tabcolsep}{0.03in}                                              
\tablenum{3}                                                                                                                                            
\tablecaption{Single-component spectral fits\label{table3}}          
\tablehead{
\colhead{} & \multicolumn{5}{c}{PL model$^a$} & \multicolumn{5}{c}{MCD model$^a$ } \\
\tableline
\colhead{Source} & 
\colhead{$\Gamma^b$} & \colhead{N$_H^f$} & \colhead{Norm} & \colhead{Good} & 
\colhead{$\chi^2$/} & 
\colhead{kT$_{in}^c$} & \colhead{N$_H^f$} & \colhead{Norm} & \colhead{Good} & 
\colhead{$\chi^2$/} \\
\colhead{} & 
\colhead{} & \colhead{} & \colhead{PL$^d$} & \colhead{fits$^h$} & 
\colhead{d.o.f.$^g$} & 
\colhead{} & \colhead{} & \colhead{MCD$^e$} & \colhead{fits$^h$} & 
\colhead{d.o.f.$^g$} \\
\colhead{} & \colhead{} &
\colhead{(10$^{21}$~cm$^{-2}$)} & \colhead{} & \colhead{} & \colhead{(keV)} & 
\colhead{(10$^{21}$~cm$^{-2}$)} & \colhead{} & \colhead{} \\
\colhead{(1)} & \colhead{(2)} & \colhead{(3)} & \colhead{(4)} & \colhead{(5)} & 
\colhead{(6)} & \colhead{(7)} & 
\colhead{(8)} & \colhead{(9)} & \colhead{(10)} &
\colhead{(11)}
}
\startdata
\multicolumn{11}{c}{\small {ULX sample}}\\
\tableline  

U1  &    2.56$^{+ 0.23}_{- 0.21}$ &     4.8$^{+  0.8}_{-  0.7}$ &  7.8$^{+12.0}_{ -1.4}\times$10$^{-5}$ &   &      68.6/53 &    0.91$^{+ 0.11}_{- 0.10}$ &     2.0$^{+  0.5}_{-  0.4}$ &  1.1$^{+ 0.8}_{ -0.4}\times$10$^{-2}$ &   &      95.5/53 \\
U2  &               2.11          &                2.4          &  5.5                                  &   &   2296.5/728 &               1.07          &                      0.7    &  6.4                                  &   &   2792.4/728 \\
U3  &    1.94$^{+ 0.14}_{- 0.13}$ &     7.6$^{+  0.9}_{-  0.8}$ &  1.4$^{+ 1.0}_{ -0.2}\times$10$^{-4}$ &   &     70.0/100 &    1.43$^{+ 0.14}_{- 0.12}$ &                 4.8$\pm$0.5 &  5.9$^{+ 2.4}_{ -1.7}\times$10$^{-3}$ & G &     91.8/100 \\
U4  &    0.81$^{+ 0.14}_{- 0.13}$ &     5.3$^{+  1.2}_{-  1.0}$ &  3.8$^{+10.3}_{ -0.6}\times$10$^{-5}$ & G &      93.1/86 &                  ($>$ 3.61) &     5.6$^{+  0.8}_{-  0.7}$ &  1.9$^{+ 2.7}_{ -0.1}\times$10$^{-4}$ & G &      94.7/86 \\
U5  &               1.70$\pm$0.05 &                 4.5$\pm$0.3 &  2.2$^{+ 0.2}_{ -0.1}\times$10$^{-3}$ & G &    304.7/289 &    1.66$^{+ 0.07}_{- 0.06}$ &                 2.5$\pm$0.2 &  6.9$^{+ 1.1}_{ -0.9}\times$10$^{-2}$ &   &    353.6/289 \\
U6  &               1.71$\pm$0.09 &                 3.7$\pm$0.5 &  6.0$^{+ 0.7}_{ -0.6}\times$10$^{-4}$ & G &    215.7/215 &    1.71$^{+ 0.13}_{- 0.12}$ &                 1.3$\pm$0.3 &  1.6$^{+ 0.5}_{ -0.4}\times$10$^{-2}$ &   &    259.8/215 \\
U7  &               2.17$\pm$0.07 &                 4.2$\pm$0.3 &  4.4$^{+ 0.8}_{ -0.3}\times$10$^{-4}$ &   &    260.3/182 &    1.13$^{+ 0.05}_{- 0.04}$ &                 2.1$\pm$0.2 &  3.9$^{+ 0.7}_{ -0.6}\times$10$^{-2}$ & G &    172.6/182 \\
U8  &    1.88$^{+ 0.16}_{- 0.15}$ &                 1.7$\pm$0.4 &  2.3$^{+ 8.1}_{ -0.3}\times$10$^{-5}$ & G &      60.8/58 &    1.21$^{+ 0.13}_{- 0.12}$ &                 0.3$\pm$0.2 &  2.1$^{+ 0.9}_{ -0.7}\times$10$^{-3}$ & G &      53.4/58 \\
U9  &    2.45$^{+ 0.22}_{- 0.20}$ &     7.0$^{+  1.3}_{-  1.2}$ &  2.0$^{+ 1.6}_{ -0.4}\times$10$^{-4}$ & G &      63.7/61 &    1.11$^{+ 0.12}_{- 0.11}$ &                 3.0$\pm$0.8 &  1.3$^{+ 0.7}_{ -0.5}\times$10$^{-2}$ &   &      76.7/61 \\
U10 &    1.00$^{+ 0.21}_{- 0.20}$ &    10.5$^{+  2.7}_{-  2.4}$ &  7.3$^{+ 3.8}_{ -1.9}\times$10$^{-4}$ & G &      72.0/81 &                     $>$3.42 &     9.3$^{+  1.2}_{-  1.1}$ &                                $<$6.8 & G &      77.4/81 \\
U11 &    1.16$^{+ 0.25}_{- 0.23}$ &    30.3$^{+  5.1}_{-  4.4}$ &  3.2$^{+ 3.0}_{ -1.0}\times$10$^{-4}$ & G &      82.8/96 &                    $>$ 3.23 &    26.6$^{+  2.3}_{-  2.1}$ &  9.6$^{+ 0.7}_{ -0.5}\times$10$^{-4}$ & G &      87.9/96 \\
U12 &    2.81$^{+ 0.71}_{- 0.63}$ &   225.7$^{+ 45.3}_{- 38.8}$ &  7.7$^{+21.1}_{ -5.3}\times$10$^{-3}$ & G &      99.8/84 &    1.89$^{+ 0.58}_{- 0.37}$ &   179.2$^{+ 30.5}_{- 26.6}$ &  2.1$^{+ 4.2}_{ -1.5}\times$10$^{-2}$ &   &     101.8/84 \\
U13 &    2.52$^{+ 0.27}_{- 0.23}$ &     2.3$^{+  0.5}_{-  0.4}$ &  4.4$^{+10.8}_{ -0.6}\times$10$^{-5}$ &   &      69.9/50 &    0.81$^{+ 0.12}_{- 0.11}$ &     0.3$^{+  0.3}_{-  0.2}$ &  1.1$^{+ 0.9}_{ -0.5}\times$10$^{-2}$ &   &     100.8/50 \\
U14 &    1.46$^{+ 0.16}_{- 0.14}$ &     5.1$^{+  1.0}_{-  0.9}$ &  6.6$^{+10.2}_{ -1.0}\times$10$^{-5}$ & G &      80.9/88 &    2.07$^{+ 0.34}_{- 0.26}$ &     3.3$^{+  0.6}_{-  0.5}$ &  1.2$^{+ 0.8}_{ -0.5}\times$10$^{-3}$ & G &      87.2/88 \\
U15 &    1.78$^{+ 0.18}_{- 0.16}$ &     6.5$^{+  1.3}_{-  1.1}$ &  7.3$^{+12.9}_{ -1.4}\times$10$^{-5}$ & G &      76.8/70 &    1.53$^{+ 0.19}_{- 0.16}$ &     4.2$^{+  0.8}_{-  0.7}$ &  2.8$^{+ 1.5}_{ -1.0}\times$10$^{-3}$ & G &      66.7/70 \\
U16 &    1.90$^{+ 0.16}_{- 0.15}$ &     3.9$^{+  0.6}_{-  0.5}$ &  5.0$^{+ 8.7}_{ -0.7}\times$10$^{-5}$ & G &      73.1/65 &    1.30$^{+ 0.14}_{- 0.12}$ &     2.1$^{+  0.4}_{-  0.3}$ &  3.2$^{+ 1.4}_{ -1.0}\times$10$^{-3}$ & G &      65.9/65 \\
U17 &               1.71$\pm$0.11 &                 7.6$\pm$0.7 &  1.5$^{+ 0.9}_{ -0.2}\times$10$^{-4}$ & G &    149.7/149 &    1.69$^{+ 0.15}_{- 0.14}$ &     5.2$^{+  0.5}_{-  0.4}$ &  4.4$^{+ 1.6}_{ -1.2}\times$10$^{-3}$ &   &    182.6/149 \\
U18 &               1.79$\pm$0.08 &     4.0$^{+  0.3}_{-  0.4}$ &          4.7$\pm$0.4$\times$10$^{-5}$ &   &    270.2/208 &    1.48$^{+ 0.10}_{- 0.09}$ &                 2.1$\pm$0.2 &  2.1$^{+ 0.5}_{ -0.4}\times$10$^{-3}$ &   &    269.4/208 \\
U19 &    1.15$^{+ 0.05}_{- 0.02}$ &                 0.6$\pm$0.1 &  2.2$^{+ 0.2}_{ -0.1}\times$10$^{-5}$ & G &    391.6/345 &    2.65$^{+ 0.21}_{- 0.22}$ &                     $<$ 0.1 &  2.6$^{+ 0.8}_{ -0.5}\times$10$^{-4}$ &   &    416.7/345 \\
U20 &    1.22$^{+ 0.13}_{- 0.11}$ &                0.2($<$ 0.5) &  1.8$^{+ 6.8}_{ -0.2}\times$10$^{-5}$ & G &      62.2/75 &    1.89$^{+ 0.26}_{- 0.24}$ &                     $<$ 0.1 &  6.1$^{+ 4.0}_{ -2.0}\times$10$^{-4}$ & G &      84.6/75 \\
U21 &    1.97$^{+ 0.11}_{- 0.10}$ &     1.4$^{+  0.3}_{-  0.1}$ &          2.8$\pm$0.2$\times$10$^{-5}$ &   &    203.7/133 &               1.16$\pm$0.08 &                     $<$ 0.2 &  2.8$^{+ 0.8}_{ -0.6}\times$10$^{-3}$ &   &    211.1/133 \\
U22 &    2.00$^{+ 0.19}_{- 0.17}$ &                 0.7$\pm$0.3 &  2.1$^{+ 7.9}_{ -0.3}\times$10$^{-5}$ & G &      48.5/53 &    0.79$^{+ 0.09}_{- 0.08}$ &                     $<$ 0.1 &  9.0$^{+ 4.3}_{ -2.9}\times$10$^{-3}$ &   &      84.3/53 \\
U23 &    1.80$^{+ 0.14}_{- 0.13}$ &                 3.4$\pm$0.5 &  1.5$^{+ 1.0}_{ -0.2}\times$10$^{-4}$ &   &      97.8/75 &    1.41$^{+ 0.14}_{- 0.12}$ &     1.7$^{+  0.4}_{-  0.3}$ &  8.3$^{+ 3.3}_{ -2.4}\times$10$^{-3}$ & G &      80.5/75 \\
U24 &    2.52$^{+ 0.28}_{- 0.25}$ &     1.9$^{+  0.5}_{-  0.4}$ &  2.1$^{+ 9.6}_{ -0.3}\times$10$^{-5}$ & G &      59.1/66 &    0.73$^{+ 0.10}_{- 0.09}$ &                 0.3$\pm$0.2 &  8.3$^{+ 5.8}_{ -3.2}\times$10$^{-3}$ & G &      70.4/66 \\
U25 &    2.25$^{+ 0.11}_{- 0.10}$ &                 1.6$\pm$0.2 &          2.5$\pm$0.2$\times$10$^{-4}$ &   &    252.3/169 &                        0.91 &                         0.1 &                                   5.0 &   &    361.5/169 \\
U26 &    1.89$^{+ 0.14}_{- 0.13}$ &                 2.5$\pm$0.4 &  2.2$^{+ 1.1}_{ -0.3}\times$10$^{-4}$ & G &      70.1/67 &    1.29$^{+ 0.13}_{- 0.11}$ &     1.0$^{+  0.3}_{-  0.2}$ &  1.6$^{+ 0.6}_{ -0.4}\times$10$^{-2}$ & G &      56.8/67 \\
U27 &               2.00$\pm$0.12 &                 3.5$\pm$0.4 &  8.3$^{+ 7.0}_{ -0.9}\times$10$^{-5}$ & G &    119.4/108 &    1.20$^{+ 0.10}_{- 0.09}$ &                 1.7$\pm$0.2 &  6.8$^{+ 2.3}_{ -1.7}\times$10$^{-3}$ &   &    134.6/108 \\
U28 &               1.88$\pm$0.12 &                 1.7$\pm$0.3 &  7.8$^{+ 7.3}_{ -0.8}\times$10$^{-5}$ & G &      82.8/83 &    1.33$^{+ 0.12}_{- 0.11}$ &                 0.3$\pm$0.2 &  5.1$^{+ 1.9}_{ -1.4}\times$10$^{-3}$ & G &      79.4/83 \\
U29 &               1.90$\pm$0.09 &     2.7$^{+  0.3}_{-  0.2}$ &  1.0$^{+ 0.6}_{ -0.1}\times$10$^{-4}$ & G &    158.7/135 &    1.28$^{+ 0.09}_{- 0.08}$ &     1.1$^{+  0.2}_{-  0.1}$ &  7.3$^{+ 2.0}_{ -1.6}\times$10$^{-3}$ &   &    204.2/135 \\
U30 &    1.26$^{+ 0.07}_{- 0.06}$ &                 0.2$\pm$0.1 &  8.6$^{+ 4.2}_{ -0.5}\times$10$^{-5}$ & G &    166.7/184 &    1.86$^{+ 0.17}_{- 0.12}$ &                     $<$ 0.1 &  3.0$^{+ 0.8}_{ -0.7}\times$10$^{-3}$ &   &    265.5/184 \\
U31 &    2.27$^{+ 0.11}_{- 0.10}$ &     5.7$^{+  0.6}_{-  0.5}$ &  2.5$^{+ 1.0}_{ -0.3}\times$10$^{-4}$ &   &    188.8/142 &               1.13$\pm$0.06 &     2.9$^{+  0.4}_{-  0.3}$ &  1.9$^{+ 0.5}_{ -0.4}\times$10$^{-2}$ & G &    154.9/142 \\
U32 &               1.79$\pm$0.11 &     6.4$^{+  0.8}_{-  0.7}$ &  1.8$^{+ 1.0}_{ -0.2}\times$10$^{-4}$ & G &    158.9/142 &    1.60$^{+ 0.13}_{- 0.12}$ &     3.9$^{+  0.5}_{-  0.4}$ &  5.6$^{+ 1.8}_{ -1.4}\times$10$^{-3}$ & G &    153.1/142 \\
U33 &    2.50$^{+ 0.14}_{- 0.13}$ &                 2.6$\pm$0.3 &  4.7$^{+ 1.1}_{ -0.5}\times$10$^{-4}$ &   &    131.5/104 &               0.80$\pm$0.06 &                 0.7$\pm$0.2 &  1.3$^{+ 0.5}_{ -0.3}\times$10$^{-1}$ &   &    179.1/104 \\
U34 &                        3.97 &                         2.9 &                                   1.7 &   &    514.9/148 &                        0.26 &                         1.5 &                                   6.3 &   &    611.5/148 \\
U35 &    2.42$^{+ 0.14}_{- 0.13}$ &                 2.6$\pm$0.5 &  2.3$^{+ 1.0}_{ -0.3}\times$10$^{-4}$ & G &     93.9/103 &               0.95$\pm$0.06 &                0.2($<$ 0.5) &  3.3$^{+ 1.0}_{ -0.8}\times$10$^{-2}$ & G &     84.2/103 \\
U36 &    3.11$^{+ 0.22}_{- 0.20}$ &                 2.0$\pm$0.3 &  3.3$^{+ 1.2}_{ -0.3}\times$10$^{-4}$ & G &      63.8/70 &               0.49$\pm$0.05 &     0.3$^{+  0.2}_{-  0.1}$ &  5.6$^{+ 3.1}_{ -1.9}\times$10$^{-1}$ &   &     105.2/70 \\
U37 &               2.43$\pm$0.12 &     1.9$^{+  0.4}_{-  0.3}$ &  1.3$^{+ 0.8}_{ -0.1}\times$10$^{-4}$ &   &    150.6/106 &               0.90$\pm$0.05 &                     $<$ 0.1 &  2.4$^{+ 0.5}_{ -0.4}\times$10$^{-2}$ &   &    154.7/106 \\
U38 &               1.86$\pm$0.09 &                 3.5$\pm$0.4 &  1.5$^{+ 0.7}_{ -0.1}\times$10$^{-4}$ & G &    178.8/151 &    1.46$^{+ 0.09}_{- 0.08}$ &     1.5$^{+  0.3}_{-  0.2}$ &  6.8$^{+ 1.7}_{ -1.3}\times$10$^{-3}$ & G &    149.1/151 \\
U39 &    2.17$^{+ 0.08}_{- 0.07}$ &                 1.9$\pm$0.1 &          1.0$\pm$0.1$\times$10$^{-4}$ &   &    430.4/282 &                        0.91 &                  0.5 &                                          2.1 &   &    800.3/282 \\
U40 &    6.51$^{+ 0.27}_{- 0.22}$ &                 3.9$\pm$0.3 &  1.9$^{+ 0.4}_{ -0.1}\times$10$^{-4}$ &   &     315.6/84 &               0.16$\pm$0.01 &                 1.3$\pm$0.1 &  1.2$^{+ 0.4}_{ -0.3}\times$10$^{ 2}$ &   &     257.2/84 \\
U41 &    3.49$^{+ 0.04}_{- 0.09}$ &     4.2$^{+  0.1}_{-  0.2}$ &          1.2$\pm$0.1$\times$10$^{-4}$ & G &    466.7/399 &               0.57$\pm$0.02 &                 0.9$\pm$0.1 &  7.4$^{+ 1.4}_{ -1.1}\times$10$^{-2}$ &   &    576.8/399 \\
U42 &    1.48$^{+ 0.17}_{- 0.16}$ &     4.7$^{+  1.6}_{-  1.4}$ &  2.1$^{+ 1.6}_{ -0.4}\times$10$^{-4}$ &   &     133.0/93 &    2.25$^{+ 0.40}_{- 0.31}$ &     2.0$^{+  1.0}_{-  0.9}$ &  2.8$^{+ 1.9}_{ -1.2}\times$10$^{-3}$ &   &     140.7/93 \\
U43 &    1.46$^{+ 0.17}_{- 0.16}$ &     5.3$^{+  1.7}_{-  1.5}$ &  1.0$^{+ 0.3}_{ -0.2}\times$10$^{-3}$ & G &      76.8/87 &    2.36$^{+ 0.42}_{- 0.31}$ &     2.4$^{+  1.1}_{-  1.0}$ &  1.2$^{+ 0.7}_{ -0.5}\times$10$^{-2}$ & G &      76.2/87 \\
U44 &    1.90$^{+ 0.26}_{- 0.24}$ &    29.6$^{+  4.4}_{-  3.9}$ &  2.0$^{+ 2.4}_{ -0.6}\times$10$^{-4}$ & G &      84.9/75 &    1.96$^{+ 0.32}_{- 0.26}$ &    22.0$^{+  2.7}_{-  2.4}$ &  2.3$^{+ 1.7}_{ -1.0}\times$10$^{-3}$ & G &      82.9/75 \\
U45 &                        2.60 &                         1.3 &                                   3.9 &   &    729.9/321 &                        0.61 &                         0.0 &                                   3.3 &   &   1436.3/321 \\
U46 &    2.29$^{+ 0.18}_{- 0.16}$ &                 2.1$\pm$0.4 &  1.1$^{+ 1.0}_{ -0.1}\times$10$^{-4}$ & G &      67.4/66 &               0.92$\pm$0.10 &                 0.3$\pm$0.2 &  2.0$^{+ 1.0}_{ -0.7}\times$10$^{-2}$ &   &      99.4/66 \\
U47 &               1.80$\pm$0.14 &                 2.3$\pm$0.4 & 10.0$^{+ 8.5}_{ -1.2}\times$10$^{-5}$ & G &      62.5/70 &    1.29$^{+ 0.14}_{- 0.12}$ &                 0.9$\pm$0.2 &  7.4$^{+ 3.3}_{ -2.3}\times$10$^{-3}$ & G &      69.9/70 \\

\tableline          
\multicolumn{11}{c}{\small Comparison sample} \\                                                   
\tableline 

C1  &    3.69$^{+ 0.10}_{- 0.09}$ &                 4.7$\pm$0.2 & 10.0$^{+ 1.1}_{ -0.6}\times$10$^{-4}$ &   &    377.3/157 &               0.55$\pm$0.02 &                 1.3$\pm$0.1 &  6.4$^{+ 1.3}_{ -1.0}\times$10$^{-1}$ &   &    561.5/157 \\
C2  &               1.49$\pm$0.05 &     1.3$^{+  0.3}_{-  0.2}$ &  8.1$^{+ 0.6}_{ -0.5}\times$10$^{-4}$ &   &    590.0/447 &               1.86$\pm$0.08 &                     $<$ 0.1 &          2.2$\pm$0.3$\times$10$^{-2}$ &   &    560.6/447 \\
C3  &               2.85$\pm$0.04 &                 2.5$\pm$0.1 &  1.5$^{+ 0.1}_{ -0.0}\times$10$^{-3}$ &   &    530.1/229 &               0.72$\pm$0.01 &                         0.3 &  5.5$^{+ 0.5}_{ -0.4}\times$10$^{-1}$ &   &    442.3/229 \\
C4  &    2.63$^{+ 0.28}_{- 0.24}$ &                 2.0$\pm$0.7 &  7.4$^{+14.0}_{ -1.3}\times$10$^{-5}$ &   &      66.9/51 &               0.74$\pm$0.07 &                     $<$ 0.2 &  2.6$^{+ 1.2}_{ -0.8}\times$10$^{-2}$ &   &      88.7/51 \\
C5  &    1.97$^{+ 0.09}_{- 0.08}$ &                 3.8$\pm$0.3 &          2.3$\pm$0.2$\times$10$^{-4}$ &   &    249.4/204 &               1.34$\pm$0.08 &                 1.7$\pm$0.2 &  1.2$^{+ 0.3}_{ -0.2}\times$10$^{-2}$ &   &    322.1/204 \\
C6  &               2.27$\pm$0.14 &     5.3$^{+  0.7}_{-  0.6}$ &  1.7$^{+ 1.0}_{ -0.2}\times$10$^{-4}$ & G &     102.9/90 &    1.09$^{+ 0.08}_{- 0.07}$ &                 2.7$\pm$0.4 &  1.5$^{+ 0.5}_{ -0.4}\times$10$^{-2}$ & G &      95.5/90 \\
C7  &               1.79$\pm$0.10 &     2.7$^{+  0.4}_{-  0.3}$ &  1.4$^{+ 0.7}_{ -0.1}\times$10$^{-4}$ & G &    139.4/126 &    1.40$^{+ 0.11}_{- 0.10}$ &                 1.1$\pm$0.2 &  7.9$^{+ 2.4}_{ -1.9}\times$10$^{-3}$ & G &    141.3/126 \\
C8  &    1.43$^{+ 0.12}_{- 0.11}$ &     1.5$^{+  0.4}_{-  0.3}$ &  6.1$^{+ 7.3}_{ -0.7}\times$10$^{-5}$ & G &      80.1/82 &    1.88$^{+ 0.25}_{- 0.20}$ &                 0.5$\pm$0.2 &  1.7$^{+ 0.8}_{ -0.6}\times$10$^{-3}$ & G &      79.9/82 \\
C9  &    2.71$^{+ 0.24}_{- 0.21}$ &                 1.9$\pm$0.4 &  8.3$^{+ 9.9}_{ -1.1}\times$10$^{-5}$ &   &      73.2/54 &               0.62$\pm$0.06 &                 0.3$\pm$0.2 &  6.0$^{+ 3.0}_{ -1.9}\times$10$^{-2}$ &   &      68.2/54 \\
C10 &    1.97$^{+ 0.17}_{- 0.16}$ &     6.8$^{+  0.9}_{-  0.8}$ &  1.7$^{+ 1.2}_{ -0.3}\times$10$^{-4}$ &   &      81.8/68 &    1.38$^{+ 0.17}_{- 0.15}$ &     4.1$^{+  0.6}_{-  0.5}$ &  7.7$^{+ 4.1}_{ -2.7}\times$10$^{-3}$ &   &     103.5/68 \\
C11 &    2.69$^{+ 0.34}_{- 0.31}$ &    42.9$^{+  6.3}_{-  5.5}$ &  5.2$^{+ 4.9}_{ -1.9}\times$10$^{-4}$ & G &      46.5/62 &    1.37$^{+ 0.18}_{- 0.15}$ &    29.5$^{+  3.7}_{-  3.3}$ &  8.6$^{+ 6.4}_{ -3.7}\times$10$^{-3}$ & G &      44.3/62 \\
C12 &    1.54$^{+ 0.14}_{- 0.13}$ &                 0.6$\pm$0.3 &  3.0$^{+ 7.5}_{ -0.3}\times$10$^{-5}$ &   &      85.3/70 &    1.40$^{+ 0.15}_{- 0.14}$ &                     $<$ 0.1 &  2.2$^{+ 0.9}_{ -0.6}\times$10$^{-3}$ &   &     100.9/70 \\
C13 &    1.81$^{+ 0.10}_{- 0.09}$ &                 0.6$\pm$0.2 &  6.2$^{+ 5.6}_{ -0.5}\times$10$^{-5}$ & G &     96.5/119 &               1.08$\pm$0.07 &                     $<$ 0.1 &  1.0$^{+ 0.3}_{ -0.2}\times$10$^{-2}$ & G &    140.9/119 \\
C14 &    2.28$^{+ 0.13}_{- 0.12}$ &                 1.0$\pm$0.2 &  7.9$^{+ 6.9}_{ -0.7}\times$10$^{-5}$ &   &    147.6/108 &               0.65$\pm$0.04 &                     $<$ 0.1 &  6.1$^{+ 1.5}_{ -1.3}\times$10$^{-2}$ &   &    233.6/108 \\
C15 &    1.46$^{+ 0.15}_{- 0.14}$ &     1.6$^{+  0.6}_{-  0.5}$ &  3.7$^{+ 9.6}_{ -0.5}\times$10$^{-5}$ & G &      54.2/62 &    1.77$^{+ 0.28}_{- 0.21}$ &     0.5$^{+  0.4}_{-  0.3}$ &  1.2$^{+ 0.7}_{ -0.5}\times$10$^{-3}$ & G &      48.2/62 \\
C16 &    1.86$^{+ 0.15}_{- 0.14}$ &                 4.7$\pm$0.9 &  1.0$^{+ 1.1}_{ -0.2}\times$10$^{-4}$ & G &      84.5/78 &    1.51$^{+ 0.16}_{- 0.14}$ &                 2.1$\pm$0.6 &  3.8$^{+ 1.7}_{ -1.2}\times$10$^{-3}$ & G &      85.6/78 \\
C17 &    1.38$^{+ 0.15}_{- 0.14}$ &     0.7$^{+  0.4}_{-  0.3}$ &  2.7$^{+ 7.3}_{ -0.3}\times$10$^{-5}$ & G &      74.2/66 &    1.81$^{+ 0.27}_{- 0.21}$ &                     $<$ 0.1 &  8.9$^{+ 5.4}_{ -3.1}\times$10$^{-4}$ &   &      83.5/66 \\
C18 &    2.60$^{+ 0.22}_{- 0.20}$ &                 2.0$\pm$0.4 &  5.5$^{+ 9.4}_{ -0.7}\times$10$^{-5}$ & G &      65.8/74 &    0.74$^{+ 0.08}_{- 0.07}$ &                0.1($<$ 0.3) &  2.0$^{+ 1.1}_{ -0.7}\times$10$^{-2}$ & G &      74.2/74 \\
C19 &    2.60$^{+ 0.17}_{- 0.16}$ &                 3.8$\pm$0.5 &  8.6$^{+ 9.3}_{ -1.2}\times$10$^{-5}$ & G &      74.5/67 &               0.87$\pm$0.07 &                 1.3$\pm$0.3 &  1.5$^{+ 0.6}_{ -0.4}\times$10$^{-2}$ & G &      65.4/67 \\
C20 &    2.35$^{+ 0.17}_{- 0.15}$ &                 3.0$\pm$0.4 &  7.5$^{+ 8.4}_{ -0.9}\times$10$^{-5}$ & G &      79.3/75 &               0.93$\pm$0.08 &     0.9$^{+  0.3}_{-  0.2}$ &  1.3$^{+ 0.5}_{ -0.4}\times$10$^{-2}$ & G &      83.0/75 \\
C21 &    2.30$^{+ 0.18}_{- 0.17}$ &     5.6$^{+  0.9}_{-  0.8}$ &  8.4$^{+11.4}_{ -1.4}\times$10$^{-5}$ &   &      85.6/63 &    1.07$^{+ 0.10}_{- 0.09}$ &     2.8$^{+  0.6}_{-  0.5}$ &  8.1$^{+ 3.4}_{ -2.4}\times$10$^{-3}$ & G &      68.8/63 \\
C22 &    5.55$^{+ 0.54}_{- 0.45}$ &     5.3$^{+  0.9}_{-  0.8}$ &  3.2$^{+ 2.2}_{ -0.5}\times$10$^{-4}$ &   &     136.9/73 &    0.26$^{+ 0.03}_{- 0.02}$ &                 1.5$\pm$0.5 &                 6.2$^{+ 6.4}_{ -3.0}$ &   &     163.0/73 \\

\enddata
\tablenotetext{a}{Model names}
\tablenotetext{b}{Photon index for the PL model}
\tablenotetext{c}{Temperature of the accretion disk at inner radius for the MCD model}
\tablenotetext{d}{Normalization constant for the PL model, in units of photons 
keV$^{-1}$~cm$^{-2}$~s$^{-1}$ at 1~keV.}
\tablenotetext{e}{Normalization constant for the MCD model, in units of 
R$_{in}$(km)$^2$~cos$\theta/$~D(10~kpc)$^2$, where
R$_{in}$(km) is the inner radius of the accretion disk in units of km, 
cos$\theta$ is the cosine of the inclination
of the accretion disk from the line of sight, and D(10~kpc) is the distance to 
the source in units of 10~kpc.}
\tablenotetext{f}{Intrinsic absorbing hydrogen column density, in units of 
10$^{21}$~cm$^{-2}$} 
\tablenotetext{g}{$\chi^2$ value for the fit and number of degrees of freedom}
\tablenotetext{h}{The ``Good'' fits, marked with a "G", have 
$\chi_{\nu}^{2}\leq~$1.2}
\end{deluxetable}


\begin{deluxetable}{l|c|cc|c|cc|c|cc|c|cc}
\rotate
\tablecolumns{12}
\tablewidth{0pt}                                                            
\tabletypesize{\tiny}   
\setlength{\tabcolsep}{0.05in}                                            
\tablenum{4}                                                                                                                                           
\tablecaption{Statistical tests for single-component fits\label{table4}}          
\tablehead{
\colhead{Samples} & 
\colhead{$\Gamma$ (PL) $^a$} & \colhead{K-S} & \colhead{T-test} &
\colhead{kT$_{in}$ (MCD) $^b$} & \colhead{K-S} & \colhead{T-test} &
\colhead{N$_H$ (PL) $^c$} & \colhead{K-S} & \colhead{T-test} &
\colhead{N$_H$ (MCD) $^d$} & \colhead{K-S} & \colhead{T-test}\\
\colhead{(1)} & 
\colhead{(2)} & \colhead{(3)} & \colhead{(4)} & 
\colhead{(5)} & \colhead{(6)} & \colhead{(7)} & 
\colhead{(8)} & \colhead{(9)} & \colhead{(10)} &
\colhead{(11)} & \colhead{(12)} & \colhead{(13)}
}
\startdata
ULX-All vs. Comp-All     &  2.11$\pm$0.89 vs. 2.36$\pm$0.92 & 0.155       & 0.324       &  1.44$\pm$0.86 vs. 1.14$\pm$0.47 & 0.457       & 0.126       &  21.51$\pm$0.50 vs. 21.42$\pm$0.42 & 0.74 & 0.49 & 21.18$\pm$0.66 vs. 21.08$\pm$0.55 & 0.84 & 0.59  \\	     
ULX-GF vs. Comp-GF    	 &  1.88$\pm$0.59 vs. 2.02$\pm$0.50 & 0.731       & 0.479       &  1.81$\pm$0.99 vs. 1.25$\pm$0.37 & 0.075       & 0.078       &  21.56$\pm$0.59 vs. 21.44$\pm$0.50 & 0.70 & 0.55 & 21.33$\pm$0.59 vs. 21.12$\pm$0.63 & 0.32 & 0.39  \\
ULX-HL vs. ULX-LL        &  1.70$\pm$0.47 vs. 2.44$\pm$1.05 & {\bf 0.011} & {\bf 0.005} &  1.89$\pm$1.02 vs. 1.12$\pm$0.50 & {\bf 0.015} & {\bf 0.001} &  21.54$\pm$0.51 vs. 21.51$\pm$0.50 & 0.75 & 0.86 & 21.25$\pm$0.70 vs. 21.13$\pm$0.64 & 0.47 & 0.58  \\
ULX-HL-GF vs. ULX-LL-GF	 &  1.52$\pm$0.39 vs. 2.18$\pm$0.57 & {\bf 0.020} & {\bf 0.001} &  2.26$\pm$1.10 vs. 1.22$\pm$0.27 & {\bf 0.022} & {\bf 0.012} &  21.58$\pm$0.60 vs. 21.54$\pm$0.61 & 0.52 & 0.87 & 21.52$\pm$0.59 vs. 21.09$\pm$0.53 & 0.43 & 0.11  \\
ULX-HL vs. Comp-All      &  1.70$\pm$0.47 vs. 2.36$\pm$0.92 & {\bf 0.021} & {\bf 0.009} &  1.89$\pm$1.02 vs. 1.14$\pm$0.47 & {\bf 0.050} & {\bf 0.003} &  21.54$\pm$0.51 vs. 21.42$\pm$0.42 & 0.72 & 0.43 & 21.25$\pm$0.70 vs. 21.08$\pm$0.55 & 0.37 & 0.45  \\
ULX-HL-GF vs. Comp-GF	 &  1.52$\pm$0.39 vs. 2.02$\pm$0.50 & 0.111       & {\bf 0.009} &  2.26$\pm$1.10 vs. 1.25$\pm$0.37 & {\bf 0.023} & {\bf 0.008} &  21.58$\pm$0.60 vs. 21.44$\pm$0.50 & 0.48 & 0.53 & 21.52$\pm$0.59 vs. 21.12$\pm$0.63 & 0.18 & 0.15  \\
ULX-LL vs. Comp-All 	 &  2.44$\pm$1.05 vs. 2.36$\pm$0.92 & 0.916       & 0.699       &  1.12$\pm$0.50 vs. 1.14$\pm$0.47 & 0.896       & 0.897       &  21.51$\pm$0.50 vs. 21.42$\pm$0.42 & 0.62 & 0.51 & 21.13$\pm$0.64 vs. 21.08$\pm$0.55 & 0.98 & 0.81  \\
ULX-LL-GF vs. Comp-GF	 &  2.18$\pm$0.50 vs. 2.02$\pm$0.50 & 0.727       & 0.460       &  1.22$\pm$0.27 vs. 1.25$\pm$0.37 & 0.637       & 0.843       &  21.54$\pm$0.61 vs. 21.44$\pm$0.50 & 0.93 & 0.64 & 21.09$\pm$0.53 vs. 21.12$\pm$0.63 & 0.97 & 0.90  \\
\enddata                                                       
\tablecomments{     
Statistical comparison using the results from single-component fits.
The samples compared in the first column are defined in section 3.1.
The abreviations are: GF, good fits, with $\chi_{\nu}^{2}\leq~$1.2;
HL, high luminosity, ULXs with X-ray (absorbed) luminosity L$_X\ge$~5.0$\times$10$^{39}$~erg~s$^{-1}$;
LL, low luminosity, ULXs with L$_X\le$~5.0$\times$10$^{39}$~erg~s$^{-1}$.
For each pair of samples in first column we performed both K-S and the T-test for the means,
and we calculate the corresponding probabilities. The significant differences, 
with probabilities $\leq~$0.05, are shown in bold.
High luminosity ULXs have harder spectra than both low-luminosity ULXs and the comparison sample.
There is also marginal evidence that ULXs show higher disk temperatures 
than the comparison sample if we only consider the good fits in both samples.}
\tablenotetext{a}{Average photon index in the PL model and 1 $\sigma$ errors}
\tablenotetext{b}{Average inner disk temperature in keV for the MCD model and one sigma errors}
\tablenotetext{c}{Average log H column density for the PL model and 1 $\sigma$ errors in units of cm$^{-2}$}
\tablenotetext{d}{The same for the MCD model}
\end{deluxetable}


\clearpage
\pagestyle{empty}  

\begin{deluxetable}{l|cccccccc}
\tablecolumns{8}   
\tablewidth{0pt}                                                                                   
\tabletypesize{\tiny}                                                                           
\tablenum{5}                                                                                       
\tablecaption{Two-component spectral fits (model PLMCD) \label{table5}}          
\tablehead{
\colhead{Source} & 
\colhead{$\Gamma^a$} & \colhead{kT$_{in}^b$} & \colhead{Norm} & \colhead{Norm} & \colhead{N$_H^e$} & \colhead{$\Delta\chi^2$/} & \colhead{Good} & \colhead{$\chi^2$/} 
\\
\colhead{} & 
\colhead{} & \colhead{} & \colhead{PL$^c$} & \colhead{MCD$^d$} & \colhead{} & \colhead{Prob.$^f$} & \colhead{fits$^g$} & \colhead{d.o.f.$^h$} 
\\
\colhead{(1)} & \colhead{(2)} & \colhead{(3)} & \colhead{(4)} & \colhead{(5)} & 
\colhead{(6)} & \colhead{(7)} & 
\colhead{(8)} &
\colhead{(9)}
}
\startdata
\multicolumn{8}{c}{\small {ULX sample}} \\
\tableline  

U1  &    2.20$^{+ 0.46}_{- 0.69}$ &    0.28$^{+ 0.17}_{- 0.12}$ &  4.9$^{+13.2}_{ -3.2}\times$10$^{-5}$ &                 1.1$^{+ 1.0}_{ -0.9}$ &     5.0$^{+  2.6}_{-  1.4}$ &         3.1/0.69 (0.69) &   &      65.5/51 \\
U2  &    2.52$^{+ 0.26}_{- 0.15}$ &    1.34$^{+ 0.03}_{- 0.06}$ &  4.0$^{+ 0.3}_{ -0.1}\times$10$^{-3}$ &                  1.4$\times$10$^{-1}$ &     2.5$^{+  0.2}_{-  0.1}$ & 397.9/$>$0.99 ($>$0.99) &   &    950.9/424 \\
U13 &    2.23$^{+ 0.38}_{- 0.36}$ &    0.13$^{+ 0.04}_{- 0.02}$ &  3.7$^{+10.3}_{ -1.3}\times$10$^{-5}$ &               $<$6.8$\times$10$^{ 3}$ &     5.0$^{+  2.0}_{-  1.6}$ &  27.4/$>$0.99 ($>$0.99) & G &      42.5/48 \\
U18 &    1.82$^{+ 0.30}_{- 0.17}$ &    0.19$^{+ 0.09}_{- 0.05}$ &  3.0$^{+10.2}_{ -0.5}\times$10$^{-5}$ &               $<$4.6$\times$10$^{ 2}$ &     6.7$^{+  4.1}_{-  2.5}$ &         8.0/0.94 (0.93) &   &      68.1/51 \\
U19 &    1.19$^{+ 0.10}_{- 0.14}$ &    0.20$^{+ 0.02}_{- 0.19}$ &  2.1$^{+ 0.2}_{ -0.3}\times$10$^{-5}$ &                 1.1$^{+ 0.9}_{ -1.0}$ &                     $<$ 4.4 &         0.8/0.09 (0.18) &   &     100.0/74 \\
U25 &    1.76$^{+ 0.28}_{- 0.30}$ &    0.19$^{+ 0.08}_{- 0.05}$ &  1.6$^{+ 1.0}_{ -0.5}\times$10$^{-4}$ &               $<$5.8$\times$10$^{ 2}$ &     2.0$^{+  1.4}_{-  0.8}$ &  12.7/$>$0.99 ($>$0.99) &   &      87.0/70 \\
U25 &    2.16$^{+ 0.19}_{- 0.21}$ &    0.13$^{+ 0.03}_{- 0.02}$ &  3.0$^{+ 1.0}_{ -0.6}\times$10$^{-4}$ &               $<$9.8$\times$10$^{ 3}$ &     3.8$^{+  1.4}_{-  1.0}$ &  33.7/$>$0.99 ($>$0.99) &   &     112.3/93 \\
U33 &    2.34$^{+ 0.23}_{- 0.37}$ &    0.20$^{+ 0.13}_{- 0.06}$ &  4.0$^{+ 1.4}_{ -1.6}\times$10$^{-4}$ &               $<$4.7$\times$10$^{ 2}$ &     2.7$^{+  1.0}_{-  0.7}$ &         3.4/0.73 (0.71) &   &    128.2/102 \\
U34 &    3.21$^{+ 0.75}_{- 0.72}$ &    0.10$^{+ 0.02}_{- 0.01}$ &  9.6$^{+22.2}_{ -4.0}\times$10$^{-5}$ &  1.4$^{+ 0.8}_{ -1.2}\times$10$^{ 5}$ &     9.0$^{+  0.8}_{-  1.1}$ &  63.6/$>$0.99 ($>$0.99) & G &      58.0/49 \\
U34 &    2.73$^{+ 0.19}_{- 0.16}$ &              0.11$\pm$ 0.01 &  2.2$^{+ 0.7}_{ -0.3}\times$10$^{-4}$ &  3.1$^{+15.0}_{ -2.4}\times$10$^{ 4}$ &     7.0$^{+  1.3}_{-  1.1}$ & 149.4/$>$0.99 ($>$0.99) &   &     130.1/93 \\
U37 &    3.35$^{+ 0.43}_{- 0.70}$ &    1.14$^{+ 0.15}_{- 0.23}$ &  1.0$^{+ 2.7}_{ -0.3}\times$10$^{-4}$ &  6.1$^{+ 5.9}_{ -2.4}\times$10$^{-3}$ &     2.4$^{+  2.7}_{-  1.4}$ &        11.2/0.98 (0.96) &   &    139.4/104 \\
U39 &    1.52$^{+ 0.23}_{- 0.29}$ &    0.25$^{+ 0.06}_{- 0.05}$ &         4.8$\pm$ 1.6$\times$10$^{-5}$ &                 3.7$^{+ 8.1}_{ -2.4}$ &     2.1$^{+  0.5}_{-  0.4}$ &  54.9/$>$0.99 ($>$0.99) & G &    142.8/139 \\
U39 &    1.28$^{+ 0.23}_{- 0.25}$ &              0.26$\pm$ 0.04 &  3.8$^{+ 5.6}_{ -1.1}\times$10$^{-5}$ &                 3.6$^{+ 5.5}_{ -2.0}$ &               2.1$\pm$  0.4 &  83.9/$>$0.99 ($>$0.99) & G &    148.5/137 \\
U40 &    3.77$^{+ 0.50}_{- 0.42}$ &              0.13$\pm$ 0.01 &  3.2$^{+ 4.9}_{ -0.9}\times$10$^{-5}$ &  4.6$^{+ 3.2}_{ -1.4}\times$10$^{ 2}$ &     1.8$^{+  0.3}_{-  0.2}$ & 169.5/$>$0.99 ($>$0.99) &   &     146.2/82 \\
U42 &                        1.82 &                        0.13 &                                   3.6 &                                   5.7 &                        11.6 &         1.6/0.42 (0.47) &   &     131.4/91 \\
U45 &    2.43$^{+ 0.12}_{- 0.10}$ &              0.12$\pm$ 0.01 &  3.5$^{+ 0.6}_{ -0.4}\times$10$^{-4}$ &  4.2$^{+ 8.9}_{ -2.4}\times$10$^{ 3}$ &     4.1$^{+  0.7}_{-  0.5}$ & 225.6/$>$0.99 ($>$0.99) & G &    212.6/182 \\
U45 &    2.28$^{+ 0.18}_{- 0.14}$ &              0.13$\pm$ 0.02 &  2.9$^{+ 0.9}_{ -0.5}\times$10$^{-4}$ &  2.8$^{+13.5}_{ -2.0}\times$10$^{ 3}$ &     4.7$^{+  1.2}_{-  0.9}$ &  96.5/$>$0.99 ($>$0.99) &   &    160.7/133 \\

\tableline                                                                             
\multicolumn{8}{c}{\small {Comparison sample}} \\                                                 
\tableline 

C1  &    3.72$^{+ 0.11}_{- 0.08}$ &              0.10$\pm$ 0.01 &         1.2$\pm$ 0.1$\times$10$^{-3}$ &  1.3$^{+ 3.3}_{ -0.9}\times$10$^{ 5}$ &               8.9$\pm$  0.4 & 137.6/$>$0.99 ($>$0.99) &   &    239.7/155 \\
C2  &    1.73$^{+ 0.13}_{- 0.18}$ &    0.12$^{+ 0.02}_{- 0.01}$ &  1.1$^{+ 0.3}_{ -0.1}\times$10$^{-3}$ &  1.3$^{+12.6}_{ -1.1}\times$10$^{ 4}$ &     4.7$^{+  1.6}_{-  2.1}$ &         8.5/0.95 (0.94) &   &    144.4/108 \\
C3  &    3.61$^{+ 0.66}_{- 0.38}$ &    0.82$^{+ 0.03}_{- 0.04}$ &  1.0$^{+ 0.2}_{ -0.1}\times$10$^{-3}$ &  2.1$^{+ 0.5}_{ -0.4}\times$10$^{-1}$ &     2.7$^{+  0.7}_{-  0.4}$ & 189.0/$>$0.99 ($>$0.99) &   &    341.1/227 \\
C4  &    2.43$^{+ 0.42}_{- 0.60}$ &    0.18$^{+ 0.17}_{- 0.02}$ &  6.1$^{+14.2}_{ -3.5}\times$10$^{-5}$ &  1.7$^{+ 2.9}_{ -1.6}\times$10$^{ 1}$ &     3.2$^{+  4.1}_{-  2.0}$ &         3.0/0.68 (0.65) &   &      63.9/49 \\
C5  &              1.96$\pm$ 0.19 &    0.20$^{+ 0.07}_{- 0.05}$ &  2.3$^{+ 1.3}_{ -0.5}\times$10$^{-4}$ &               $<$7.0$\times$10$^{ 2}$ &     6.4$^{+  2.2}_{-  1.6}$ &     12.8/$>$0.99 (0.98) & G &    164.7/139 \\
C9  &               2.54($<$3.49) &    0.59$^{+ 0.06}_{- 0.14}$ &  2.8$^{+ 9.0}_{ -2.8}\times$10$^{-5}$ &  5.2$^{+11.6}_{ -1.2}\times$10$^{-2}$ &     1.1$^{+  1.4}_{-  0.9}$ &         6.8/0.92 (0.90) &   &      66.4/52 \\
C10 &    1.99$^{+ 0.17}_{- 0.19}$ &              0.18($<$ 0.74) &  1.8$^{+ 0.4}_{ -0.3}\times$10$^{-4}$ &               $<$8.5$\times$10$^{ 1}$ &               9.7$\pm$  0.6 &         2.2/0.59 (0.60) &   &      79.6/66 \\
C12 &    1.42$^{+ 0.23}_{- 0.51}$ &                        0.31 &  2.5$^{+ 7.5}_{ -1.5}\times$10$^{-5}$ &               $<$1.4$\times$10$^{-1}$ &                0.6($<$ 1.3) &         1.2/0.39 (0.39) &   &      84.0/68 \\
C14 &    1.54$^{+ 0.37}_{- 0.52}$ &    0.35$^{+ 0.08}_{- 0.07}$ &  3.0$^{+ 6.9}_{ -1.6}\times$10$^{-5}$ &  4.3$^{+ 5.6}_{ -2.3}\times$10$^{-1}$ &     0.5$^{+  0.4}_{-  0.3}$ &  17.4/$>$0.99 ($>$0.99) &   &    130.2/106 \\
C22 &    3.49$^{+ 0.62}_{- 0.56}$ &              0.12$\pm$ 0.02 &  8.9$^{+36.8}_{ -3.6}\times$10$^{-5}$ &  2.2$^{+20.7}_{ -1.7}\times$10$^{ 3}$ &     5.3$^{+  1.7}_{-  1.1}$ &  57.4/$>$0.99 ($>$0.99) & G &      79.6/71 \\

\enddata  
                                                         
\tablenotetext{a}{Photon index for the PL model}
\tablenotetext{b}{Temperature of the accretion disk at the inner radius for the MCD model in keV}
\tablenotetext{c}{Normalization constant for the PL model as in Table~3}
\tablenotetext{d}{Normalization constant for the MCD model, as in Table~3}
\tablenotetext{e}{Intrinsic absorbing H column density in units of 10$^{21}$~cm$^{-2}$}
\tablenotetext{f}{F-test $\Delta\chi^2$/ confidence levels for the model PLMCD against the PL model alone.
The values in parentheses are obtained from simulations; see Section~3.2 for details.}
\tablenotetext{g}{The ``good'' fits are marked with a ``G'' as in Table~3}
\tablenotetext{h}{The $\chi^2$ value for the fit and number of degrees of freedom}

\end{deluxetable}

\clearpage
\pagestyle{plaintop}

\begin{figure}
\rotate
\epsscale{0.95}
\plotone{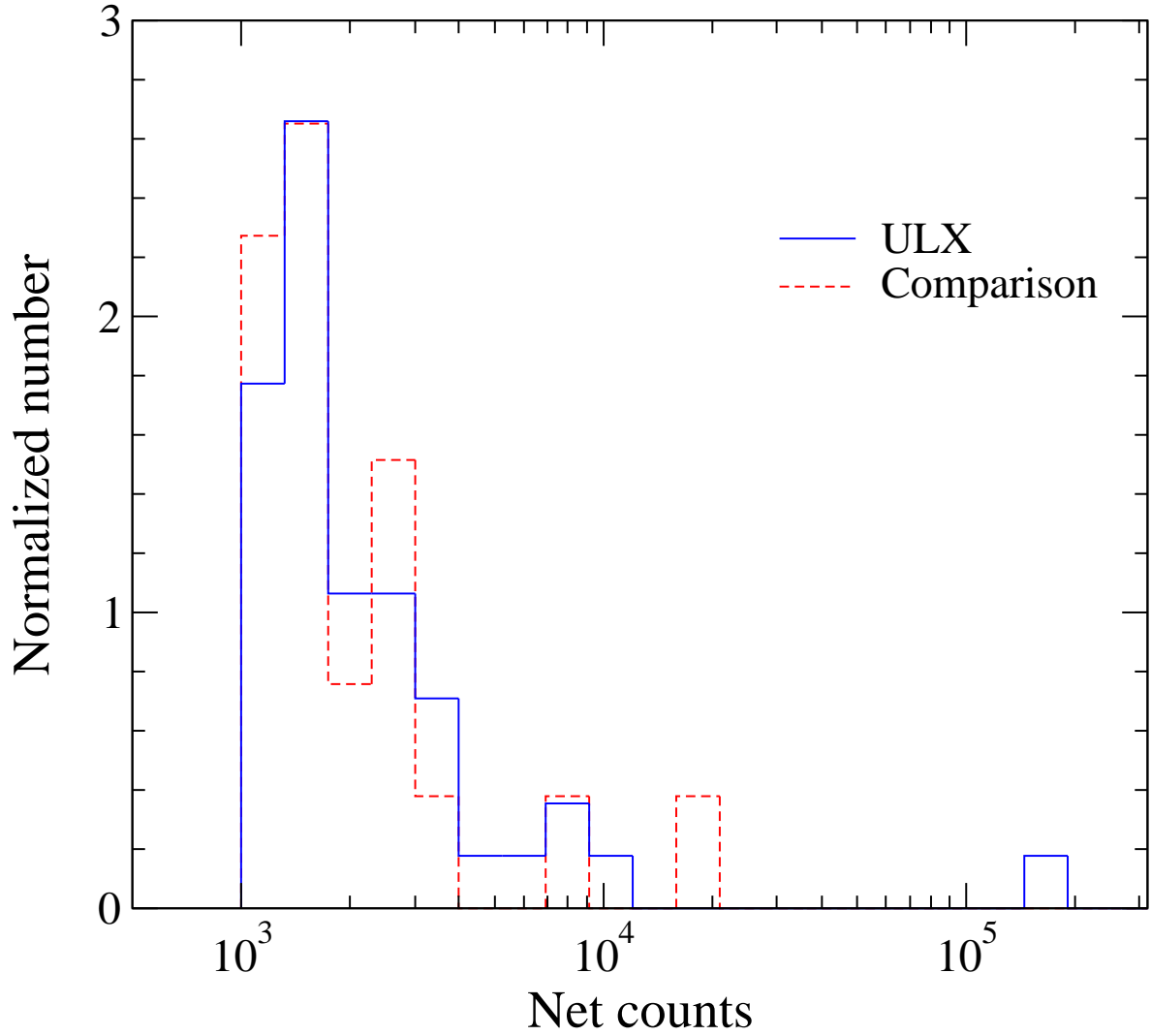}
\caption{
Normalized histograms of net counts for the ULX and comparison samples.
For multiple observations we use the highest number of counts for each object.
The histograms are normalized to unit area.
The data with counts $>$10$^5$ are from one source: the long observation of M33~X-8 (U2).
}
\end{figure}

\begin{figure}
\plottwo{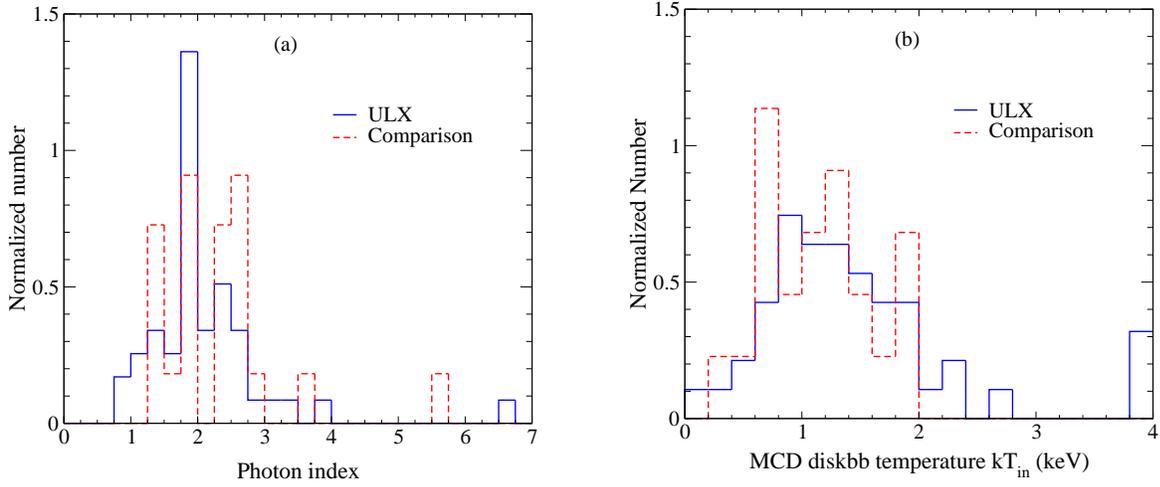}{f2b.eps}
\caption{
Normalized histograms from single-component fits.
The histograms are normalized to have a unit area.
(a) Photon index distribution from PL fits.
(b) Inner disk temperature distribution from MCD fits.
}
\end{figure}

\begin{figure}
\rotate
\plottwo{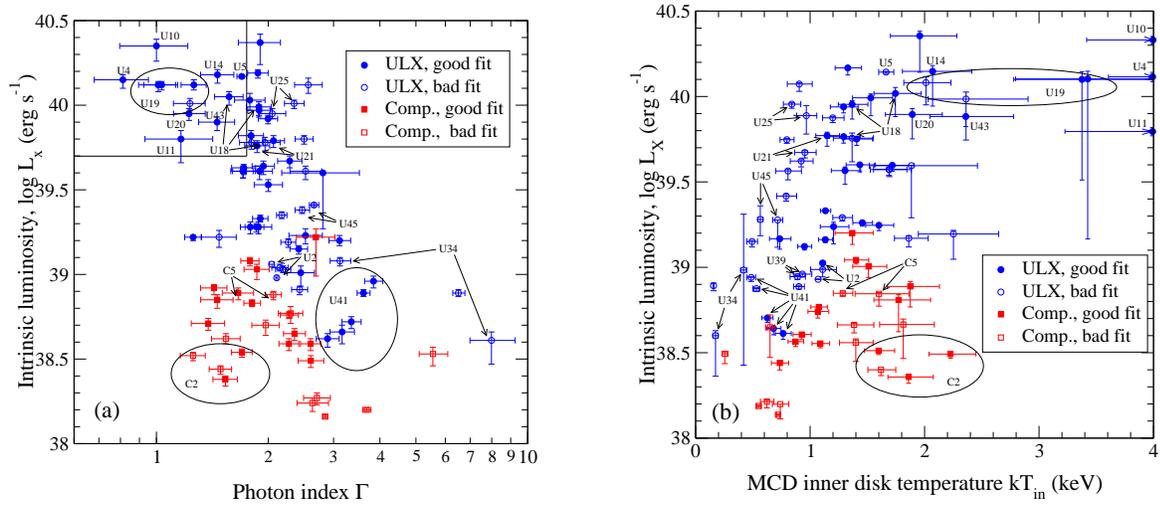}{f3b.eps}
\caption{
(a) Luminosity vs. photon index from PL model fits. 
(b) Luminosity vs. inner disk temperature using the MCD model.
In the upper left corner of the left panel we define a subsample of 9 unique ULXs
(U19 has 4 observations). They have luminosities $>$5$\times$10$^{39}$~erg~s$^{-1}$ 
and $\Gamma<$~1.7. For clarity, we label only the objects in this ULX subsample plus any 
objects with multiple observations.
}
\end{figure}

\begin{figure}
\plottwo{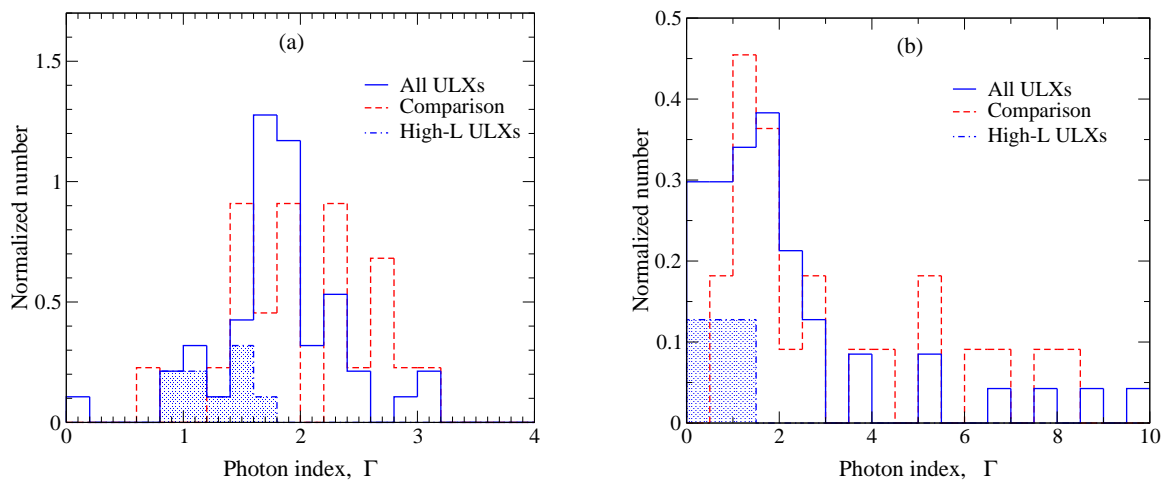}{f4b.eps}
\caption{
Histograms for photon indices from spectral fits with fixed inner disk temperatures,
for ULX and lower luminosity samples, both normalized to unit area for easy comparison.
We also show the high-luminosity, hard ULXs (filled blue regions).
No significant difference is seen between ULXs and the comparison sample,
but the high-luminosity ULXs are distinctly harder (i.e., flatter spectra).
a) Model PLMCD0.25.
b) Model PLMCD1.0.
}
\end{figure}

\begin{figure}
\rotate
\plotone{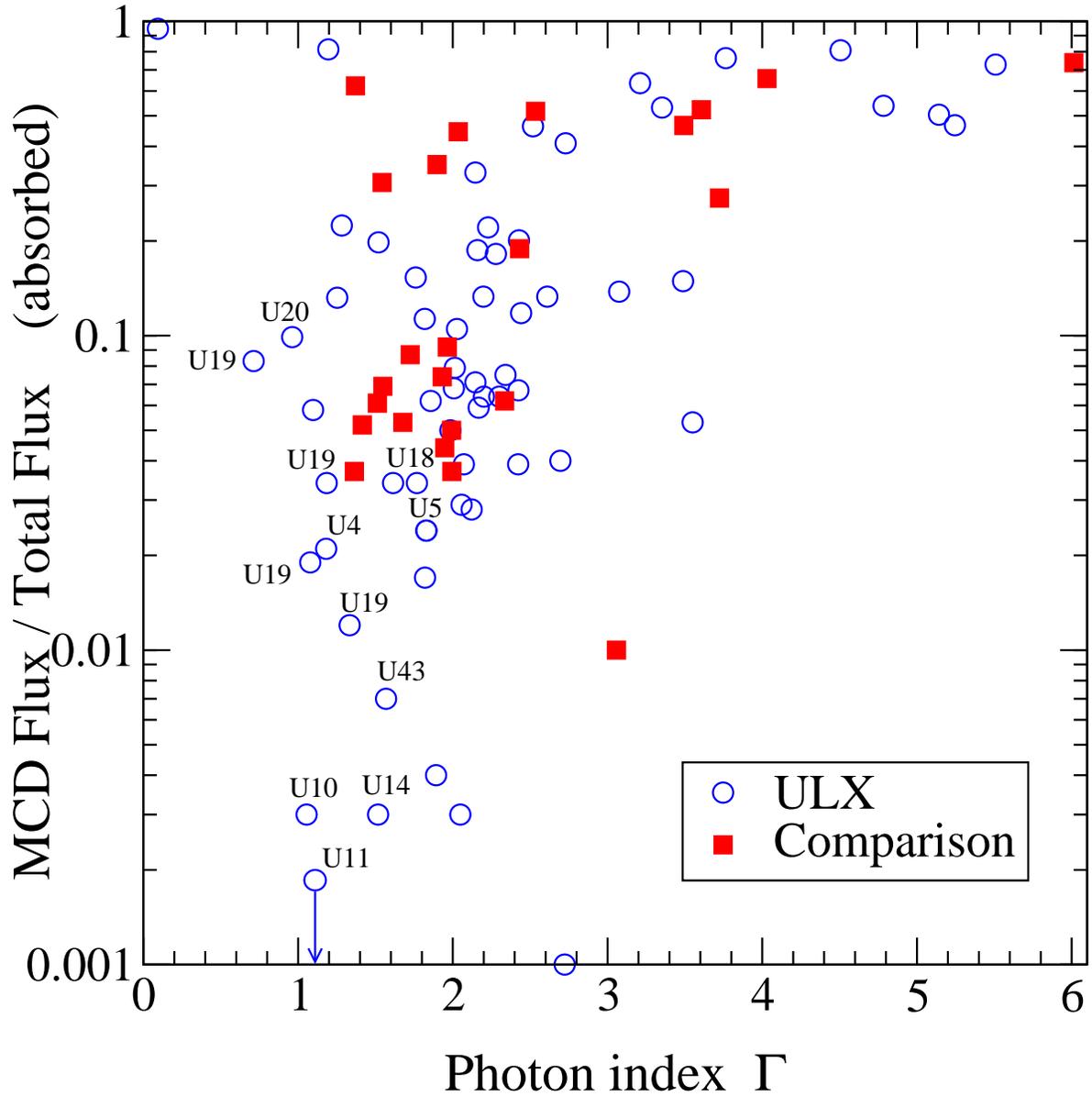}
\caption{
Ratio of MCD blackbody flux to the total flux (MCD fraction), plotted  against photon index,
using the free parameter model PLMCD. The fluxes are absorbed.
For clarity, we only label the 9 high-luminosity, hard ULXs as defined in Figure~3a.
These have both the hardest spectra and the lowest flux contribution from the MCD components.
For U11 the fraction is below 0.001. 
}
\end{figure}

\begin{figure}
\rotate
\plottwo{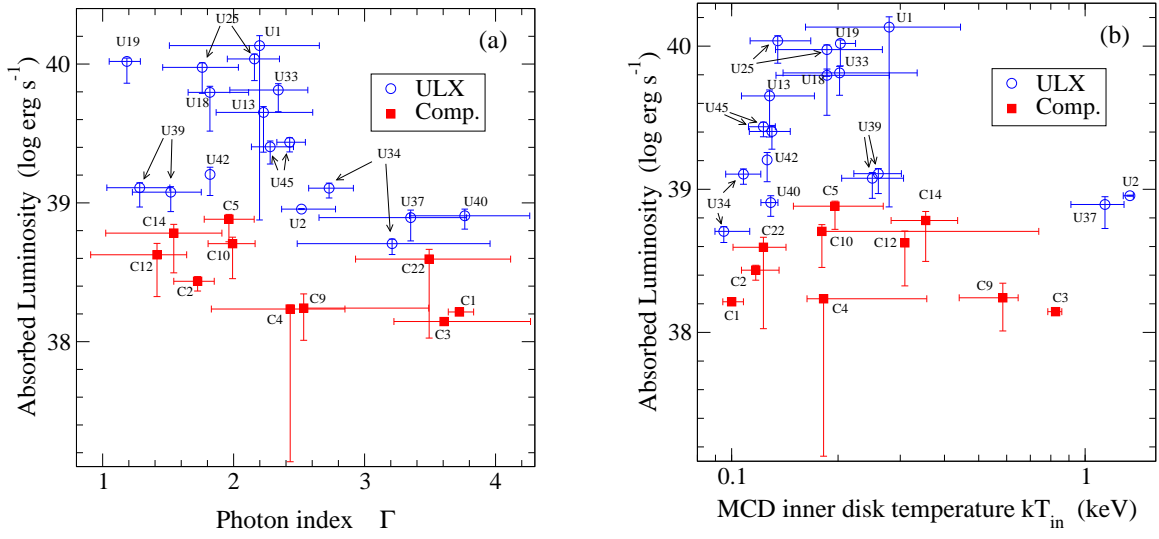}{f6b.eps}
\caption{ 
Absorbed luminosity scatter plots from the two-component spectral model with free parameters (PLMCD).
We present results only for the spectra that did not provide acceptable fits with single component models.
a) Photon index dependence.
b) Disk temperature dependence.
}
\end{figure}

\end{document}